\documentclass[a4paper,12pt,DIV=12]{scrartcl}
\pdfoutput=1
\usepackage{amsmath,amssymb}
\usepackage{color,xcolor}
\definecolor{blue}{rgb}{0,0,0.5} 
\usepackage[colorlinks,linkcolor=blue,urlcolor=blue,citecolor=blue]{hyperref}
\usepackage{graphicx}
\addtokomafont{disposition}{\rmfamily\boldmath}
\usepackage[utf8]{inputenc}
\usepackage{enumerate} 
\usepackage{slashed}
\usepackage{cite} 
\usepackage{comment}
\usepackage{multirow}
 \usepackage{bbold}
\allowdisplaybreaks
\usepackage{bm,amssymb,slashed,graphicx,multirow,soul,mathtools,xspace,array}  
\usepackage{float}                   
\allowdisplaybreaks
\usepackage{bbold}
\usepackage{subfigure}
 \usepackage{slashed}
 \usepackage{acronym}

\DeclareOldFontCommand{\tt}{\normalfont\ttfamily}{\mathtt}

\newcommand{\al}{\alpha}

\newcommand{\ga}{\gamma}

\newcommand{\la}{\lambda}

\newcommand{\TeV}{\,{\textrm TeV}}
\newcommand{\GeV}{\,{\textrm{GeV} }}

\newcommand{\matel}[3]{\langle #1|#2|#3\rangle}

\newcommand{\mi}{\!-\!}

\newcommand{\SEC}{Sec.~}

\newcommand{\APP}{App.~}
\newcommand{\APPs}{Apps.~}

\definecolor{violet}{rgb}{0.94, 0.2, 0.8}
\definecolor{lightblue}{rgb}{0.39, 0.58, 0.93} 
\definecolor{asparagus}{rgb}{0.53, 0.66, 0.42}

\begin{document}


\begin{center}
{\Large\bfseries \boldmath $A_{\textrm{CP}}[D_{(s)}^{0,+} \to V  \gamma]$ from Large  ${\cal O}_8$}\\[0.8 cm]
{\large%
James Lyon
and Roman Zwicky,
\\[0.5 cm]
\small
Higgs Centre for Theoretical Physics, School of Physics \& Astronomy, University of Edinburgh, 
Peter Guthrie Tait Road, King's Buildings, Edinburgh EH9 3FD, Scotland, UK
} \\[0.5 cm]
\small
E-Mail:
\texttt{\href{mailto: jamesly0n@hotmail.com}{ jamesly0n@hotmail.com}},
\texttt{\href{mailto:roman.zwicky@ed.ac.uk}{roman.zwicky@ed.ac.uk}}.
\end{center}

\abstract{
CP-violation in  $\Delta A_{\textrm {CP}} = -0.154(29)$, 
in the  $D^0 \to \pi^+\pi^-/K^+K^-$ system,  is established and its central value is 
one  order of magnitude above the naive Standard Model (SM) estimate.
It remains unclear whether this is due to  currently   
incalculable strong interaction matrix elements or genuine new physics (NP) such as a shift in 
${\cal O}_8$ with a weak phase.
We show that  interference of the long-distance (LD) terms with the ${\cal O}_8$ matrix element
can give rise to $A_{\text{CP}}^{D \to V \gamma} = \textrm{few}  \cdot 10^{-3}$ 
(for reference values $ {\textrm{Im}}[C_8^{NP}] \approx 10^{-3}$). 
In addition, it is pointed  out that the ratio of  left- to right-handed (photon polarisation) 
LD amplitudes is  measurable in  time-dependent CP (TDCP) asymmetries. 
We argue that both theory and experimental consideration favour  weak annihilation (WA) 
as the dominant LD contribution. 
More definite progress could be achieved by  either computing the radiative corrections to WA 
or the measurement of the charged modes $D^+_{(d,s)} \to  (\rho,K^{*})^+ \gamma$  and $D_s \to \rho^+ \gamma$.
}

\setcounter{footnote}{0}

\tableofcontents

\section{Introduction}
\label{sec:charmCP}

\subsection{$A_{\textrm{CP}}$ in $D^0 \to \pi\pi/KK$}

CP-violation is parametrically  suppressed in the charm sector (of order ${\cal O}(10^{-4})$).
In 2011  LHCb  \cite{LHCb:2011osy} and CDF \cite{CDF:2011ejf} reported a value of 
CP-violation in the hadronic system $D^0 \to \pi^+\pi^-/K^+K^-$, 
$\Delta A_{CP}  = -0.65(18)$ with central value 
considerably above  expectation. Since then CP-violation in the charm system 
has been established \cite{LHCb:2019hro,HFLAV:2019otj}
\begin{equation}
\label{eq:CPex}
\Delta A_{CP} = A^{K^+K^-}_{\textrm{CP}} - A^{\pi^+\pi^-}_{\textrm{CP}} = -0.154(29)  \cdot  10^{-2} \;.
\end{equation}
at a lower central value. {However,}  the question of whether this is NP or  due to hadronic matrix elements considerably  above its naive expectation remains unclear and is part of the 
investigation of this paper. 
In \eqref{eq:CPex},  $A_{\textrm{CP}}^f$ is
\begin{equation}
A_{\textrm{CP}}^f \equiv 
\frac{\Gamma[D^0 \to f] - \Gamma[\bar D^0 \to f] }{\Gamma[D^0 \to f] + \Gamma[\bar D^0 \to f] } \;,
\end{equation}
a shorthand for the time integrated  CP-asymmetry, for a case where the final state $f$ is a CP-eigenstate. 
$\Delta A_{CP}$ is a  convenient quantity since
systematic experimental errors cancel. 
It is worthwhile to add that if $SU(3)_F$, or more precisely $U$-spin, were a good symmetry then 
$A^{K^+K^-}_{\textrm{CP}} = -A^{\pi^+\pi^-}_{\textrm{CP}}$.  
In the quantity $\Delta A_{\textrm{CP}}$ the  TDCP-asymmetry part cancels. Effects can remain through  time-acceptance differences in the $\pi$- and $K$-system.
 Although the latter is  estimated to be small, e.g. \cite{LHCb:2011osy}. 
Hence  direct (i.e. time-independent CP-asymmetry) is expected to be responsible for the relatively large value of $\Delta A_{CP}$.

Sizeable direct CP-asymmetries  necessitate  
large  strong (CP-even) and weak  (CP-odd) phase differences in two amplitudes of comparable size
(cf. App.~\ref{app:CPbasic}). 
The reason CP-violation is believed to be small in the charm system is that 
the weak phases are suppressed by four powers of the Cabibbo angle 
(or Wolfenstein parameter $\la \approx 0.23 $),  leading to the naive expectation 
$\Delta A_{\textrm{CP}} \approx \text{few} \cdot 10^{-4}$.
 In the non-leptonic case the QCD matrix elements, which determine the strong phase as well as
the ratio of amplitudes, are difficult to compute from first principles as the size of the charm mass is neither
suited to chiral nor heavy quark theory. Advances in lattice QCD open the door to first principles results \cite{Hansen:2012tf} and should be available in the foreseeable future.
Thus the question of whether the large central value \eqref{eq:CPex}, should it remain, is due to 
NP \cite{Bigi:2011re,Rozanov:2011gj,Cheng:2012wr,Li:2012cfa} or somewhat unexpected strong dynamics 
\cite{Brod:2011re,Bhattacharya:2012ah,Feldmann:2012js,Franco:2012ck}, such as in the $\Delta I = 1/2$-rule 
$K \to \pi\pi$ system\footnote{It was pointed out quite some time ago  \cite{Golden:1989qx} that an enhancement 
of the triplet transition, in the SU(3)-flavour classification, may lead to sizeable CP-violation.
 E.g. $A^{PP}_{\textrm{CP}} \approx 0.08\cdot 10^{-2}$ which would lead to $|\Delta A_{\textrm{CP}}| \approx 0.16  \cdot 10^{-2}$ which is not far off the central value in \eqref{eq:CPex}.}, is an open question at present.

Taking the viewpoint that the asymmetry is largely due to NP it turns out 
that  a weak  phase in the $|\Delta C| = 1$ chromomagnetic operator\footnote{Note 
that this is the sign convention 
of \cite{Isidori:2011qw} but opposite to references  \cite{Giudice:2012qq,Isidori:2012yx}.}
\begin{equation}
\label{eq:O8}
 {\cal O}_8   \equiv -  \frac{ g m_c   }{8 \pi^2} \,  \bar u  \sigma\! \cdot \! G  (1+\gamma_5) c   \;, \quad {\cal O}'_8  \equiv  -  \frac{ g m_c   }{8 \pi^2} \,  \bar u  \sigma\! \cdot \! G  (1-\gamma_5) c \;,
 \end{equation}
($ \sigma  \cdot  G =  \sigma_{\mu\nu}   G_a^{\mu\nu}\lambda^a/2$), appears to be a promising candidate \cite{Isidori:2011qw}; not contradicting observations
 such as $D^0$-$\bar D^0$-mixing. 
Note that the ${\cal O}_8^{(')}$-operators  are of the $\Delta I = 1/2$-type and do not fall 
into the testable $\Delta I = 3/2$-class \cite{Grossman:2012eb}.
Furthermore, ${\cal O}_8'$ is the structure which 
is the less abundant helicity in the SM due left-handedness of the weak interactions; $[C'_8/C_8]|_{\textrm{SM}} \approx m_u/m_c$.

To get an idea of the size of the NP contribution \cite{Giudice:2012qq} 
one might resort to naive factorisation (NF) e.g. \cite{Grossman:2006jg}. Slightly extending the notation in \cite{Giudice:2012qq} 
one gets,
\begin{equation}
\label{eq:aCPC8}
\Delta A_{CP}^{\textrm{NP}}|_{\textrm{NF}} \approx -1.8 \left( {\textrm{Im}}[C_8^{NP}]    - 
{\textrm{Im}}[ C_8^{'NP}] \right) \sin \delta     \;,
\end{equation}
where $\delta$ is the unknown  strong phase difference between the  $KK$ and $\pi\pi$ rescattering states which 
is expected to be sizeable.
Note that since the sign of $\sin \delta$ is unknown  the $D^0 \to \pi\pi/KK$-system  there is additional ambiguity on the $C_8^{(')} $ Wilson coefficients.
Since the decay of a  $J^P(D^0)  = 0^-$ particle into  two  $J^P(\pi/K) = 0^-$ particles necessitates 
parity violation 
only the $\gamma_5$-part in \eqref{eq:O8} contributes and therefore results in  opposite signs 
of ${\textrm{Im}}[C_8^{NP}]$ and ${\textrm{Im}}[C_8^{'NP}]$ in \eqref{eq:aCPC8} respectively.
Now,  a value of
\begin{equation}
\label{eq:NC8}
\left( {\textrm{Im}}[C_8^{NP}]    - 
{\textrm{Im}}[ C_8^{'NP}] \right) \sin \delta  \;, \approx  10^{-3} \qquad \text{naive factorisation (NF)} \;,
\end{equation}
 could account for the central number
in \eqref{eq:CPex}.   
One has to bear in mind that \eqref{eq:NC8} is due to 
NF and could easily be out by    factors of a few.
We   take 
\begin{equation}
\label{eq:ref}
{\textrm{Im}}[C_8^{(')NP}]  =  10^{-3}\;,
\end{equation}  
as our reference value which  is consistent with    \cite{Giudice:2012qq}  after adjusting to the current experimental value  \eqref{eq:CPex}.
This value is at least   two to three orders of magnitude above the SM-value for $C_8$, 
cf. \APP~\ref{app:Ceff}, and  additionally suppressed by  $m_u/m_c$ 
$C_8'$.

  NP models that could induce such values as in \eqref{eq:NC8}
without violating existing constraints  are  
 supersymmetric models \cite{Grossman:2006jg,Giudice:2012qq,Hiller:2012wf}, {leptoquarks  \cite{deBoer:2017que}}, 
 Randall-Sundrum flavour anarchy \cite{Delaunay:2012cz} and models of partial compositeness \cite{Keren-Zur:2012buf}, whereas 
 in fourth family models it seems more difficult to accommodate \cite{Feldmann:2012js}.

\subsection{$A_{\textrm{CP}}$ in $D^0 \to V \gamma$}

The question of whether  $C_8$-values like \eqref{eq:ref} lead to  observable effects elsewhere, or specifically in 
$D \to V \ga$, is the subject of this paper.
 It was pointed out in reference  \cite{Isidori:2012yx} that sizeable direct CP-violation 
in  $D^0 \to (\rho^0,\omega) \gamma$ can be induced through  ${\textrm{Im}}[C_7]$ provided that 
the LD amplitude carries a strong phase.\footnote{Other channels and effects that were proposed are the electric dipole moment of the nucleon \cite{Giudice:2012qq,Mannel:2012hb}, 
CP-asymmetry in $D^0 \to \phi \to K^+K^-$ \cite{Fajfer:2012nr} 
 and  $D^0 \to V(\to PP) \to K^+K^-$  \cite{Cappiello:2012vg}.} 
The latter is necessary as the short distance (SD) contribution of ${\cal O}_7$
does not come with  a strong phase.  Let us emphasise a few points that are either new or improved in our paper as compared with the literature:
\begin{itemize}
\item With regards to $A_{\textrm{CP}}$ in $D^0 \to V \gamma$ and \cite{Isidori:2012yx} our discussion
includes the ${\cal O}_8$ matrix element  which carries a strong phase. 
Thus  for (direct) CP-violation no sizeable  LD phase  is required. 
\item We observe that WA is the dominant LD mechanism since it is enhanced  over quark loops 
(QL) by two loop factors.  In the neutral modes this hierarchy is weakened by the colour suppression of WA and we indeed 
find that in practice $|{\cal A}_{QL} |/ |A_{WA}|_{D^0}$ could be close to $30\%$ (cf. App.~\ref{app:WAvsQL}).
\item We provide partial radiative corrections in WA in terms of the $D \to \ga$ form factor, in \SEC\ref{sec:amp} is a new result of this paper. 
\item In order to overcome the colour suppression, which manifests itself in large scale uncertainties,  
we emphasise the need for the computation of the \emph{full} radiative corrections for the neutral modes.
We motivate the experimental measurement of the charged modes for which the colour suppression is not present in practice. 
\item We observe that  TDCP is solely sensitive to long-distance contributions and that its 
long-distance chirality is measurable in the 
neutral modes (cf. \SEC\ref{sec:timeCP} and also \cite{deBoer:2017que,Adolph:2018hde} for further elaborations).
\end{itemize}

 \vspace{0.2cm}

The paper is organised as follows.
In \SEC\ref{sec:HeffAmp}  notation is introduced and the basics of CP-violation, specific to 
the charm sector, is reviewed.  
\SEC\ref{sec:main} is the main part of this paper: 
the  amplitudes  are detailed and estimates for   direct  and time
dependent CP-violation are given 
(using the matrix elements of the operator ${\cal O}_8^{(')}$  \cite{Dimou:2012un}).
Conclusions and discussions are presented in \SEC\ref{sec:epilogue}.
An important part of our work is the discussion  of the LD contribution reported in  \APP\ref{app:long}. Furthermore  \APPs \ref{app:CPbasic} 
and \ref{app:comment}  contain further material on CP-violation in general and specific to the decay in question. 

\section{Effective Hamiltonian and Amplitudes}
\label{sec:HeffAmp}

\subsection{$|\Delta C| = 1$ Hamiltonitan}
 \label{sec:HC=1}
 
 Following, closely, the notation of \cite{Isidori:2011qw} we write the  effective $\Delta C =1$ SM Hamiltonian as follows 
 \begin{equation}
 \label{eq:Heff_generic}
 {\cal H}^{\textrm{eff}} = \lambda_d  {\cal H}_d +  \lambda_s  {\cal H}_s +  \lambda_b  {\cal H}_{\textrm{peng}}
 \;, \quad \lambda_{D} \equiv V_{cD}^* V_{uD} \;, D =d,s,b
 \end{equation}
 and
   \begin{eqnarray}
  \label{eq:Heff}
  {\cal H}_q &=&  \frac{G_F}{\sqrt{2}} \sum_{i=1}^2 C_i^q {\cal O}_i^q + {\textrm{h.c.}}  \;,  \qquad q = d,s \nonumber 
  \\[0.1cm]
   {\cal O}_1^q &=&  (\bar u L_\mu q) (\bar q L^\mu c)       \;, \qquad
 {\cal O}_2^q =  (\bar u_\alpha L_\mu q_\beta) (\bar q_\beta L^\mu c_\alpha)  \nonumber 
  \\[0.1cm]
  \lambda_b {\cal H}_{\textrm{peng}} &=&  \frac{G_F}{\sqrt 2} \left( C_7 {\cal O}_7 + C'_7 {\cal O}'_7  + C_8 {\cal O}_8 + C'_8 {\cal O}'_8  + ... \right)
 \end{eqnarray}
 with $L_\mu \equiv  \gamma_\mu  (1-\gamma_5)$ and $\alpha,\beta$ being colour indices.
 The Hamiltonian  ${\cal H}_{\textrm{peng}}$ contains all the SD transitions 
 including electric \eqref{eq:O7} and chromomagnetic \eqref{eq:O8}  operators as well
 as the four quark operators with structure different from  ${\cal O}_{1,2}$.
 As compared to  \cite{Isidori:2011qw} we have absorbed the  $\lambda_b$ into the Wilson coefficient which is non-standard for the SM contribution.
Since  $\lambda_{d,s}  = {\cal O}( \lambda)$ and 
$\lambda_{b} =  {\cal O}( \lambda^5)$, where $\lambda \approx 0.226$ \cite{ParticleDataGroup:2020ssz} is the Wolfenstein parameter, 
one gets using the unitarity relation 
\begin{equation}
\label{eq:unitarity}
 \lambda_d + \lambda_s + \lambda_b =  0 \;, \quad \Rightarrow \quad 
 \lambda_d \approx - \lambda_s  \;, \quad  \lambda_b \approx  0 \;,
\end{equation}
where the symbol $\approx$ above is to be understood as up to corrections of   ${\cal O}(\lambda^4)$. 
The fact that the third generation  decouples up to ${\cal O}(\lambda^4)$ is the reason why in the SM 
the generic expectation for CP-violation 
is  $A_{\textrm{CP}} \approx \text{few} \cdot {\cal O}(\lambda^4)$ as mentioned in the introduction. 

 \subsection{Parametrisation of decay rate}
 \label{sec:Drwgamma}


We write the amplitude as follows\footnote{The amplitudes ${\cal A}_{\perp,\parallel}$ up to phases are often denoted by ${\cal A}_{\textrm{PC,PV}}$ in the literature e.g. \cite{Burdman:1995te,Isidori:2012yx}. The acronyms PC and PV stand for parity-conserving and -violating respectively.}
\begin{alignat}{2}
\label{eq:A1A2}
& {\cal A}[D \to  V \gamma] \equiv  \matel{V\gamma}{{\cal H}^{\textrm{eff}}}{D} &\;=\;&  {\cal A}_\perp \frac{P_\perp}{2}   + 
  {\cal A}_\parallel  \frac{P_\parallel}{2}   \nonumber \\[0.1cm]
 & &\;=\;& {\cal A}_L    \left( \frac{P_\perp+P_\parallel}{4} \right) + {\cal A}_R  \left( \frac{P_\perp- P_\parallel}{4} \right) \;,
\end{alignat}
with $P_\perp  =  2 \epsilon^{}_{\rho  \alpha \beta \gamma} 
\epsilon^{*\rho}   \eta^{*\alpha} p^{\beta}q^\gamma$ and
$P_\parallel = 2 i \{(p \!\cdot q) ( \eta^{*} \! \cdot \! \epsilon^{*})  \mi 
(\eta^*\!\cdot\! q) (p \!\cdot\! \epsilon^{*})  \}$
where 
 $\eta(p)$ and $\epsilon(q)$ are the vector meson and photon polarisation tensors and
the Levi-Civita convention is settled   by ${\textrm{tr}}[\gamma_5 \gamma_a \gamma_b \gamma_c \gamma_d] = 4i \epsilon_{abcd}$. 
It is noted that ${\cal A}_{L(R)} \equiv ({\cal A}_\perp \pm {\cal A}_\parallel)$
  correspond to left- and right-handed
polarised photons.
 The rate \cite{Burdman:1995te}, in our conventions, is given by 
\begin{equation}
\label{eq:rate}
\Gamma[D \to V\gamma] = \frac{1}{32 \pi} m_D^3 \left(1-\frac{m_V^2}{m_D^2} \right)^3 \left( |{\cal A}_\perp|^2 + |{\cal A}_\parallel|^2  \right) \;.
\end{equation}

\section{CP-asymmetries  in $D \to V \gamma$}
\label{sec:main}

The operators \eqref{eq:O8} consist of $c \to u$ transitions 
of the FCNC type. In a heavy-to-light transition for which LCSR can make predictions \cite{Dimou:2012un}
the $c$-quark can pair with a $u$, $d$ or $s$-quark. 
This leads to the following  possible transitions with CP-violation  
$D^0 \to (\rho^0,\omega) \gamma$, $D^+ \to  \rho^+ \gamma$ and $D_s^+ \to  K^{*+}  \gamma$. 
The transitions $D^0 \to \bar K^{*0} \ga$ and $D_s^+ \to \rho^+ \ga$    are not of the FCNC type and do not lead to CP-violation in our 
framework (cf. Tab.~\ref{tab:decays}  {in App.~\ref{app:overview}} for more info and benchmark values for the rates). 
 Note that it is only for the neutral $D^0$-system that oscillations
and thus TDCP asymmetries are feasible.

As previously mentioned and outlined in App.~\ref{app:CPbasic} direct CP-violation 
originates in its minimal form by two amplitudes with  weak and strong phase difference.
In this work, these two amplitudes  are the LD and the NP-enhanced ${\cal O}_8$ contributions respectively. 

\subsection{Weak annihilation and $O_8$ amplitudes}
\label{sec:amp}

\begin{figure}[ht]
 \centerline{\includegraphics[width=2.0in]{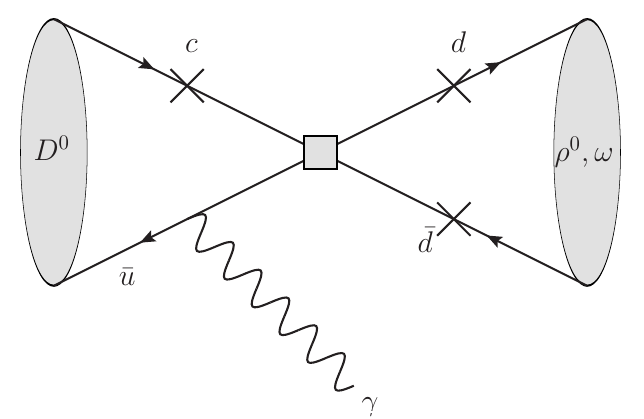}
 \includegraphics[width=2.0in]{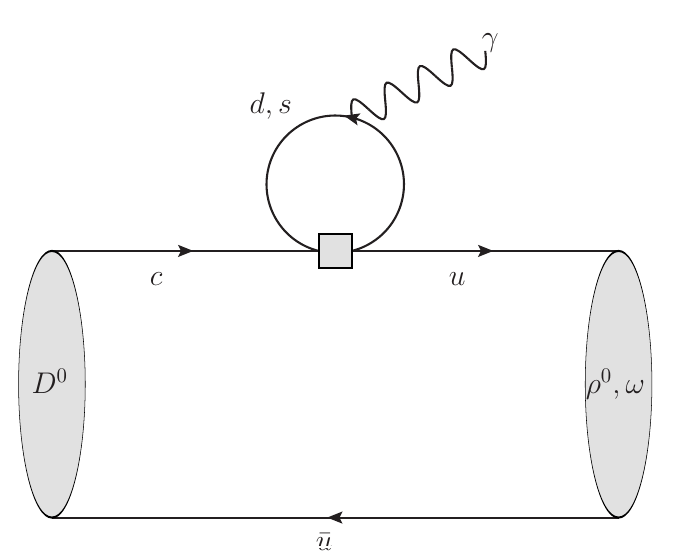}  \includegraphics[width=2.0in]{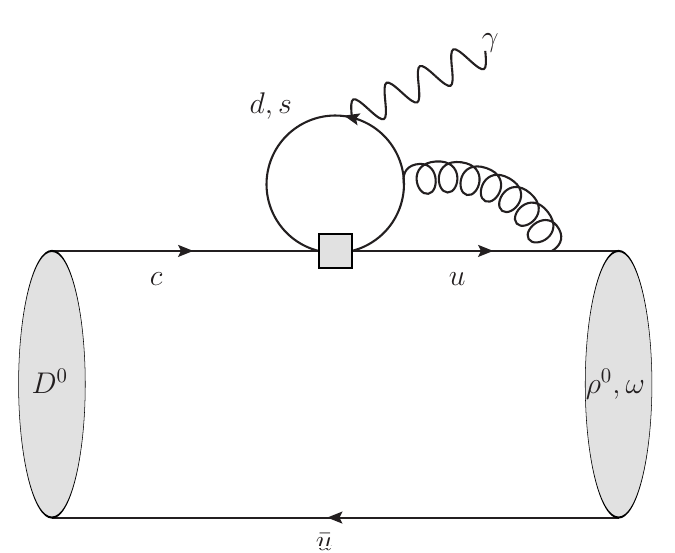} }
 \caption{\small A selection of LD diagrams for $D^0 \to (\rho^0,\omega) \gamma$. Note that it is the fact that the 
 $\rho^0/\omega$ carry both $\bar dd $ and $\bar uu$ components that makes the same operator ${\cal O}^{d,s}_{2}$ \eqref{eq:Heff} contribute to both (WA \&  QL)  topologies.
 (left) weak annihilation (WA). (centre) quark loop (QL).  This contribution vanishes, exactly, for on-shell
 photon, by virtue of gauge invariance as discussed in the text. (right) QL example of ${\cal O}(\alpha_s)$-correction. This diagram has a sizeable imaginary part which can be inferred from the computation for 
 $c \to u \gamma$ in reference \cite{Greub:1996wn}. }
 \label{fig:graphs}
 \end{figure}

The  WA contribution is extensively  discussed in  
\APP\ref{app:long} for which we give an executive summary here. Firstly, it is argued that 
 WA  dominates over the QL (cf. Fig.~\ref{fig:graphs}  (left) and 
(centre,right)) from a theoretical and experimental viewpoint in \APPs\ref{app:WAvsQL} and
\ref{app:exWA} respectively.  
In  \APP\ref{app:LCSR} we elaborate  
on making concrete predictions in the neutral modes $D^0 \to V \ga$, which concern an unfortunate 
cancellation of Wilson coefficients and highlights the need (for currently) unavailable 
 radiative corrections. 
 This is followed in \APP\ref{app:overview} by an overview and comparison 
of all $D^{0,+} \to V \ga$ modes.  The  main and important conclusion of  \APP\ref{app:LCSR} 
is  that  the situation can  considerably be improved by i)  the measurement of the charged modes and or ii) the computation of the radiative correction to WA.

Here we discuss aspects  related to CP-violation.  
The WA contribution comes with a small weak phase and the strong phase  should not be sizeable either. 
The former  is a  direct consequence of the CKM-hierarchy. The amplitude is
 proportional to $\lambda_{d,s} \approx  {\cal O}(\lambda)$ and its weak phase is of the order of ${\cal O}(\lambda^4)$. 
 The strong phase is small in the sense 
that it originates from radiative corrections to the WA diagram  (e.g. Fig.~\ref{fig:graphs} (right)).

Restricting ourselves to the LO contribution, as NLO is currently not available, the amplitude 
is given by ($X = \perp,\parallel$)\footnote{The factor $c_V$ is inserted to absorb trivial factors
due to the wave function decomposition $\rho^0(\omega) \sim \frac{1}{\sqrt{2}} ( \bar uu \mp \bar dd)$. $c_V= - \sqrt{2}$ for $\rho^0$ in $c \to d$, $c_V = \sqrt{2}$ in 
all other transitions into $\omega$ and  $\rho^0$ and $c_V=1$ otherwise. Note that in the overall CP-asymmetry this
factor will drop out.}
\begin{equation}
\label{eq:AX}
{\cal A}_X(D^{0(-)} \to V^{0(+)} \ga) = \kappa_{0(+)} \frac{e G_F}{\sqrt{2} c_V} \la_{\textrm CKM} a_{2(1)} \frac{m_V f_V}{m_D} V_X^{D^{0(+)} \to \ga}(m_V^2) \;, \quad  
\end{equation}
in the convention of \cite{Janowski:2021yvz}, adapting $D_\mu = \partial_\mu - i e A_\mu$ for the sign of the covariant derivative in this work, $\la_{\textrm CKM} $ is the product of the appropriate CKM factors and for the decay constants we use the values given in App.~C of \cite{Bharucha:2015bzk}.
The factor  
$\kappa_{0(+)} = 2(1)$ is the empirically motivated scaling factor  (cf. \APP\ref{app:LCSR}).
The functions $V_{\perp,\parallel}$ are the $D \to \ga$ form factors which are the only contribution at leading order in the chiral limit in the SM.
We use  the NLO form factor results  \cite{Janowski:2021yvz}, evaluating them 
we   obtain 
 the new results
\begin{alignat}{4}
\label{eq:ALD}
& V^{D_0 \to \ga}_\perp(m_\rho^2) &\;\approx \;& -0.55      \;,\quad   & & 
 V^{D_0 \to \ga}_\parallel(m_\rho^2) &\;=\;&  -0.17   \;, \nonumber \\[0.1cm] 
& V^{D^+ \to \ga}_\perp(m_\rho^2) &\;=\;&  +0.067    \;,\quad   & & 
 V^{D^+ \to \ga}_\parallel(m_\rho^2) &\;=\;& +0.35 \;,   \nonumber \\[0.1cm] 
& V^{D_s \to \ga}_\perp(m_{\rho}^2) &\;\approx \;& +0.11    \;,\quad   & & 
 V^{D_s \to \ga}_\parallel(m_{\rho}^2) &\;=\;&  +0.44  \;.
\end{alignat}
{The continuum threshold of $s_0  = 6 \GeV^2$ is well between $(m_D + 2 m_\pi)^2 \approx 4.6\GeV^2$ and $(m_D + m_\rho )^2 \approx6.9\GeV^2$.  The Borel parameter is chosen as a compromise value to render  the (partonic) OPE convergent 
and to suppress the continuum contributions in the hadronic contribution. }
We refrain from an uncertainty analysis as we only aim for rough estimates in order to motivate experimental searches. Moreover, 
$V^{D^+ \to \ga}_\perp(m_\rho^2) \approx V^{D^+ \to \ga}_\perp(m_\omega^2) $  and 
$V^{D_s \to \ga}_\perp(m_\rho^2) \approx V^{D_s \to \ga}_\perp(m_{K^*}^2) $ hold to sufficient precision.
  The reference values $a_2 = C_2 + C_1/3 \approx - 0.5$ and $a_1 = C_1+C_2/3  \approx 1$ correspond to the 
colour suppressed and colour allowed combination of Wilson coefficients (cf.    App.~\ref{app:overview} and  \cite{Buras:1994ij} for further discussion).

We now turn to the ${\cal O}_8$-contribution of 
  the chromomagnetic operator \eqref{eq:O8}. Its amplitude  is parametrised as follows,
\begin{equation}
\label{eq:AO8}
{\cal A}_{i}|_{8} =
  \matel{V\gamma}{{\cal H}^{\textrm eff}|_8 }{D}   
  =   \frac{G_F}{\sqrt{2}} \left( \frac{e m_c }{2 \pi^2} \right)   \, \frac{1}{c_V}\, 
\begin{cases}  (C_8 +C_8')G_1(0)  \quad i = 1  \\ (C_8 - C_8') G_2(0) \quad   i = 2\end{cases}   \;,
\end{equation}
where ${\cal H}^{\textrm eff}|_8 = \frac{G_F}{\sqrt 2} ( C_8 {\cal O}_8 + C'_8 {\cal O}'_8)$.
Therefore  $G_{1,2}(0)$  corresponds to the matrix elements, with on-shell photon $q^2 =0$,
\begin{equation}
\label{eq:MO8}
\matel{V\gamma}{{\cal O}^{(')}_8}{D} = \left( \frac{e m_c }{4 \pi^2} \right) \frac{1}{c_V} (G_1(0) P_\perp \pm G_2(0) P_\parallel) \;,
\end{equation}
 analogous to the penguin matrix element for $T_1$ and $T_2$ Eq.~\eqref{eq:MO7} and  $e =  \sqrt{ 4 \pi \alpha} > 0$ is the electromagnetic charge. 
In our notation $G_{1(2)} = G_{\perp(\parallel)}$ but refrain to do so.
Moreover, 
$G_1^{D^0 \to \rho^0\gamma}(0) \approx G_1^{D^0 \to \omega \gamma}(0)$, 
$G_1^{D_s^+ \to K^{*+}\gamma}(0) \approx G_1^{D^+ \to \rho^+ \gamma}(0)$  to sufficient  accuracy 
for our purposes and $G_1(0) = G_2(0)$ holds {at twist-$2$} accuracy   \cite{Dimou:2012un} which we employ for our estimates.\footnote{In fact the ratio 
of the WA to the $G_1(0)$ form  factor is well approximated by $R= r_\rho/ r_\omega$ where 
$r_X = ( f^\perp_X)/(m_X f^\parallel_X) $ is the ratio of the tensor to the vector decay constant. Information on this ratio exists only sparsely in the literature. Similar remarks apply to the $D_s^+ \to K^{*+}$ and $D^+ \to \rho^+$-transitions.}  In particular the imaginary part, relevant for the CP-asymmetry,
is found to be 
\begin{equation}
\label{eq:ImG1}
{\textrm Im} [ G_1^{D_0}(0)]  \approx  - 0.20(8) \;,  \quad {\textrm Im} [ G^{D^+}_1(0)]  \approx  - 0.10(4) \;,
\end{equation}
where numbers were rounded.
The values in \eqref{eq:ImG1} are sizeable 
compared to typical estimates $T^{D^0}_1(0) \approx T_1^{D^+}(0) \approx 0.7$ of the-${\cal O}_7$ operator 
as compiled in  \cite{Isidori:2012yx}.
The difference in the numerical value of neutral and charged matrix elements in Eq.~\eqref{eq:ImG1} 
originate from different  charges of the valence quarks of the mesons.
Using the reference value for ${\textrm Im}[C_8^{(')}]$ the relevant ratios are around 
\begin{equation}
\label{eq:rAMP}
\frac{ | {\cal A}_{\perp,\parallel}| _{8}}{{\cal A}_{\perp,\parallel}|_{\textrm LD}|} =  \textrm{few}  \cdot 10^{-3} \;,
\end{equation}
  and thus the scale for direct CP-violation is set at the sub-percent level for the reference value \eqref{eq:ref}.

\subsection{Direct CP-violation}
\label{sec:directCP}

Since the photon polarisation is not easy to measure in practice  a slightly inclusive rate 
$\Gamma[D \to V \gamma] = \Gamma[D \to V \gamma_L] + \Gamma[D \to V \gamma_R]$ is measured.
We parametrise the corresponding amplitudes as follows,
\begin{equation}
\label{eq:ALR}
{\cal A}_{L,R} = {\cal A}_\perp \pm  {\cal A}_\parallel  =    l_{L,R} e^{i \delta_{L,R}} + g_{L,R} e^{i \Delta_{L,R}} e^{i \phi_{L,R}}
\end{equation}
with 
\begin{alignat}{2}
\label{eq:detailsLR}
& l_{L(R)} &\;=\;&   |l_\perp \pm l_\parallel|  \;, \quad l_{\perp,\parallel} \equiv  {\cal A}_{\perp,\parallel}|_{\textrm{LD}}  \nonumber \\
& g_{L(R)} e^{i \Delta_{L(R)}}  &=\;&  \frac{G_F}{\sqrt{2}} \left( \frac{e m_c }{2 \pi^2} \right)   \frac{1 }{c_V}  \,  |C_8^{(')}|  2 {G_{L,R}} (0) \nonumber \\
& {G_{L,R}(0)} &=\;& { |G_{1,2}(0)|e^{i \Delta_{L,R}}} \;, \quad C_8 = |C_8|e^{i \phi_L} \, \quad C'_8 = |C'_8|e^{i \phi_R} \;,
\end{alignat}
where $\Delta_{L,R}$, $\delta_{L,R}$  and $\phi_{L,R}$ are the strong and the weak phase of \eqref{eq:AO8} respectively leaving the 
quantities $l_{L(R)},g_{L(R)}$  real-valued. 
In the equation above we have made
use of $G_1(0) = G_2(0)$, found at leading twist \cite{Dimou:2012un},  implying that ${\cal O}_8$  and
${\cal O}_8'$ solely  contribute to the left- and right-handed amplitude respectively and in addition 
leads to $\Delta_L = \Delta_R$. 
The latter is not true when the contribution due to ${\textrm{Im}}[C_7^{(')}]$ is included, in which case
the formulae for $g_{L,R}$ have to be  modified according  to Eq.~\eqref{eq:tg} in App.~\ref{app:time}.

In the case where the two photon polarisations are not distinguished the formula for 
CP-violation is slightly more complicated than the one given in Eq.~\eqref{eq:ACP}. 
The general formulae and a 
derivation, including TDCP-asymmetries, can be found in the \APP of  Ref.~\cite{Muheim:2008vu} for example.  Using the corresponding standard formulae 
for the amplitude \eqref{eq:ALR} yields
\begin{eqnarray}
A_{\textrm{CP}}(D^0 &\to& V\gamma) = \frac{- 4 }{n}   \left( g_L l_L \sin(\Delta_L - \delta_L)   \sin(\phi_L) +   \{ L \leftrightarrow R \}  \right)  \;, \\[0.1cm]
n  &\equiv& 2( l_L^2 + 2  \left( g_L l_L \cos(\Delta_L- \delta_L)  \cos(\phi_L)   g_L^2 \right) +   \{ L \leftrightarrow R \}  )  \;. \nonumber 
\end{eqnarray}
Assuming $l_{L(R)} \gg  g_{L(R)}$ and imposing $\Delta \equiv  \Delta_L = \Delta_R$ one gets
\begin{eqnarray}
\label{eq:ACPapprox}
A_{\textrm{CP}}(D^0 \to V\gamma) &\approx&    \frac{ - 2 } {l_L^2+  l_R^2 }  \left( g_L l_L \sin(\Delta- \delta_L)   \sin(\phi_L) +   \{ L \leftrightarrow R \}  \right)  \;.
 \end{eqnarray}
In the absence of a computation, and in view of the chiral suppression at leading order, we  set
the LD phases $\delta_{L,R}$ \eqref{eq:ALR} to zero in the remaining formulae but it will be taken into account in the error budget.
This allows us to express  $A_{\textrm{CP}}$ in terms of the quantities discussed at the beginning of the paper
 \begin{eqnarray*}
A_{\textrm{CP}}(D^0 \to V\gamma)  &=&  \frac{ - 4  } {l_L^2+  l_R^2 }  \frac{G_F}{\sqrt{2}} \left( \frac{e m_c }{2 \pi^2} \right)  \frac{{\textrm{Im}}[G_1(0)]}{c_V}  \left(  l_L {\textrm{Im}}[C_8]  +   l_R  {\textrm{Im}}[C_8']   \right)  \;.
\end{eqnarray*}
This formula, modulo notation,  reduces to $A_{\textrm{CP}}$ \eqref{eq:ACP} for $l_\perp = l_\parallel$ (i.e. $l_R = 0$).

With $m_c = 1.3\GeV$, Eqs.~\eqref{eq:NC8}, \eqref{eq:ALD} and \eqref{eq:ImG1} 
we get for the neutral transitions
\begin{eqnarray}
A_{\textrm{CP} }(D^0 \to (\rho^0,\omega) \gamma)   
&=&     \pm ( 3.0 {\textrm{Im}}[C_8^{NP}]  
 + 1.6  {\textrm{Im}}[C_8^{'NP}]  )   
\label{eq:ACPexplicit}
\end{eqnarray}
where the difference between $\rho_0$ and $\omega$ due to mass and decay constants is negligible
compared to the estimated   uncertainty of about $50\%$ (to be discussed further below).
In going from \eqref{eq:ACPapprox} to \eqref{eq:ACPexplicit} we have used the fact that the 
imaginary part of $C_8^{\textrm{SM}}$, which contains the CKM prefactors, is negligible with respect to the
values \eqref{eq:NC8}. For the charged transitions we get
  \begin{eqnarray}
A_{\textrm{CP} }(D^+_{(d,s)} \to  (\rho,K^{*})^+ \gamma)   
 =  (4.6,3.3)   {\textrm{Im}}[C_8^{NP}]  
-   (3.1,2.0) {\textrm{Im}}[C_8^{'NP}]  
\label{eq:ACPexplicit2}
\end{eqnarray}  
where we recall  our reference value ${\textrm{Im}}[C_8^{(')NP}]  = 10^{-3}$ \eqref{eq:ref}. 
Again the  uncertainty is estimated to be about $50\%$.
Note that the different sensitivity of 
$\Delta A_{\textrm{CP}}$, $A_{\textrm{CP} }(D^0 \to (\rho^0,\omega) \gamma) $ and  $A_{\textrm{CP} }(D^+_{(d,s)}\to (\rho^+,K^{*+}) \gamma$ with respect to ${\textrm{Im}}[C_8]$ and ${\textrm{Im}}[C_8']$
gives a  handle to discriminate between the individual contributions 
of the two chromomagnetic operators.  
 
 Let us turn to the discussion of the uncertainty.
   The major uncertainty comes from the estimate of the ${\cal O}_8$
  matrix elements which we estimate to be around $35\%$ \cite{Dimou:2012un}. Then there is the phase of the WA contribution, $\delta_{L,R}$, for which we assign an uncertainty $ |\delta_{L,R}| = 45^\circ$ {based on the estimate that the radiative corrections of WA could be of equal size as leading order with maximal $90^\circ$-phase. Note that more than $90\%$ degrees itself is again unrealistic since the rate suggests that the interference is not destructive. 
 In summary, this  leads to an uncertainty of approximately  $30\%$.}  
Amongst the LD contributions the combination 
   $l_L^2 + l_R^2$ is taken from experiment but the ratio $l_L/l_R$ which we took from 
   \cite{Janowski:2021yvz} could have uncertainties, say, at the $20\%$-level. 
    Adding the three   sources discussed above in quadrature, as they would seem uncorrelated, 
  we get about $50\%$ uncertainty. A few additional remarks are in order.
  In App.~\ref{app:ACPSM} we estimate the SM contribution to be of   the order of  $10^{-4}$
  which is negligible.
  Furthermore we refrain from including at this point the uncertainty due to the $C_7$-effect discussed in App.~\ref{app:ACP87}.
 We would like to mention though that  it cannot be excluded, depending on the model and 
 the LD phase, 
 that the $C_7$ and $C_8$-effect  conspire to cancel  significantly in the CP-asymmetry.

\subsection{Time-dependent CP-violation}
\label{sec:timeCP}

As a result of $D^0$-$\bar D^0$ oscillations CP-asymmetries are time dependent for the neutral meson,
giving rise to novel features.
In particular TDCP asymmetries do not necessitate a strong phase difference in the two amplitudes. 
Thus in principle we have to adjust the amplitudes to include the $C_7$-effect, from $g_{L,R}$ 
 Eqs.~(\ref{eq:ALR},\ref{eq:detailsLR})  to $\tilde g_{L,R}$ \eqref{eq:tg}  as detailed in App.~\ref{app:time}.
 Indications are though that these effects are overshadowed  by the dominance 
 of the  LD amplitudes $l_{L,R}$.

Important mixing parameters of the $D^0$-$\bar D^0$  system are the mass and width difference, the mixing phase $\phi_{D}$  as well as the ratio $|p/q|$ of the parameters $p$ and $q$ translating between the 
flavour and mass eigenstates.
The latest HFAG values \cite{HFLAV:2019otj} are 
\begin{alignat}{2}
\label{eq:mixing}
& x_D = \frac{\Delta m_D}{\Gamma}  = 0.409(48) \cdot 10^{-2}  \;, \qquad & \left|\frac{p}{q} \right|_D  \approx& 1 \;, \nonumber \\[0.1cm]
 & y_D = \frac{\Delta \Gamma_D}{2 \Gamma}  = 0.719(113) \cdot 10^{-2} \;,  &\phi_{D} \approx& -13(13)(4)^\circ
 [ -6(11)(4)^\circ] \;,
\end{alignat}
where $\Gamma = (\tau_{D^0})^{-1}$ is the inverse lifetime of the $D^0$-mesons and $\Delta m_D$ 
and $\Delta \Gamma_D$ are the difference of the heavy and the light $D^0$-meson mass and width respectively. 
Above we did set the value for $|p/q|=1$ as both the no direct CP allowed and direct CP allowed   value are compatible with $1$ 
within very small uncertainties. The value for  $\phi_{D}$ we quote both values no direct CP allowed and direct CP allowed (in brackets). 
 Assumeing $|p/q|_D =1$, 
the TDCP-asymmetry assumes the following form
\begin{equation}
\label{eq:CPform}
 A_{\textrm{CP}}(D  \to V  \gamma)[t]
= \frac{S \sin(\Delta m_D t) -C \cos(\Delta m_D t)}
{ {\textrm{cosh}}(\frac{\Delta \Gamma_D}{2}t) - H  {\textrm{sinh}}(\frac{\Delta \Gamma_D}{2} t) }    \;, 
\end{equation}
where the convention  $A_{\textrm{CP}}(0) = -C$ is somewhat awkward but standard.  
The formulae for $S$ and $H$ are given in App.~\ref{app:time} and $C$ from the previous section.
Let us define the LD chirality-asymmetry (ratio) by
\begin{equation}
\label{eq:chiLD}
\chi_{\textrm{LD}} \equiv \frac{l_\perp^2-l_\parallel^2}{l_\perp^2 + l_\parallel^2} = \frac{2  l_L l_R}{l_L^2 + l_R^2}  \in [-1,1]\;.
\end{equation}
With   values as in \eqref{eq:ALD} we get $\chi_{\textrm{LD}} \approx 0.8(1)$.
Thus if we assume $\chi_{\textrm{LD}}   \gg 10^{-2}$, $l_{L,R}  \gg \tilde g_{L,R}$, which both seem true, 
and once more set  $\delta_{{L,R}}=0$ 
we get an interesting expression for for $H$ and $S$,
\begin{eqnarray}
\label{eq:LDchirality}
H[S]  \approx  \frac{2  l_L l_R}{l_L^2 + l_R^2} \cdot  \left( - \xi \cos[\sin](\phi_{D}) \right) =  \chi_{\textrm{LD}} \cdot \left(  - \xi \cos[\sin](\phi_{D}) \right) \;,
\end{eqnarray}
which directly measures the ratio of the LD chirality structure times the cosine and sine of the mixing 
angle of the $D^0$-system. The variable $\xi = \pm 1$ is the CP-eigenvalue of the $V$-meson whose
values can be found in App.~\ref{app:time}.  With $\xi(\rho^0,\omega)= 1$ we get
\begin{alignat}{2}
& H[D^0 \to (\rho^0,\omega)\gamma]  &\;\approx\;&  -0.8(1) \cos (\phi_{D})   \;, \nonumber  \\[0.1cm]
&   S[D^0 \to (\rho^0,\omega)\gamma]   &\;\approx\;&  -0.8(1) \sin (\phi_{D})  \;.
\end{alignat}
Let us emphasise once more that this relation is valid in the case where a left- and right-handed 
amplitude are  comparable in size and dominate all the other contributions. 

The experimental tractability of $S$ and/or
 $H$  depends on the angle $\phi_D$. 
Should $\phi_D$ \eqref{eq:mixing},
that is to say $\sin \phi_D$, 
turn out to be sizeable then $S$ could be measured as for $B \to K^* \gamma$ at the B-factories. 
If $\cos \phi_D$ is sizeable, which is what the value in  \eqref{eq:mixing} indicates, then one would need
to focus on $H$. The latter might be measured, in analogy to $B_s \to \phi \gamma$ case \cite{Muheim:2008vu}, 
  in the rates $D^0 \to (\rho^0,\omega) \gamma$ and the one for $\bar D^0$ 
  without flavour-tagging, which has experimental advantages, though it has to be added that 
  the relatively small width difference in the $D^0$ system, $y_D/y_{B_s} \approx 0.1$, means that 
 roughly a hundred times more data has to be accumulated to achieve the same precision on $H$ 
 in the $D^0$- as in the $B_s$-system. 
 We further refer the reader to the works \cite{deBoer:2017que} and \cite{Adolph:2018hde} where some of these ideas have been extended 
 to baryon decays $\Lambda_c \to p \gamma$ and $1^+$ final state mesons  ($D^0 \to K_1 \ga$)  respectively. The $1^+$-modes combined with the $1^-$-modes 
 have the potential to discriminate between LD and SD per se \cite{Gratrex:2018gmm}.

\section{Discussion and Conclusions}
\label{sec:epilogue}
 
 Partly building up on ideas in \cite{Isidori:2012yx} we have shown how ${\textrm{Im}}[C_8]$ 
 and ${\textrm{Im}}[C_8^{'}]$ become observable in CP-asymmetries 
 in $D \to V \gamma$.  Setting the LD phases $\delta_{L,R} =0$ \eqref{eq:ALR}, in the absence of a computation, we got
 \eqref{eq:ACPexplicit} and \eqref{eq:ACPexplicit2}
\begin{alignat}{2}
\label{eq:summary}
& A_{\textrm{CP} }(D^0 \to (\rho^0,\omega) \gamma)   
 &\; \approx\; &     \pm  (3.0 {\textrm{Im}}[C_8^{NP}]  
+ 1.6  {\textrm{Im}}[C_8^{'NP}] )       \;, \nonumber
\\[0.1cm] 
& A_{\textrm{CP} }(D^+_{(d,s)} \to  (\rho,K^{*})^+ \gamma)   
 &\; \approx\; &  (4.6,3.3)   {\textrm{Im}}[C_8^{NP}]  
-   (3.1,2.0) {\textrm{Im}}[C_8^{'NP}]  \;,
\end{alignat}
where we recall  our reference values ${\textrm{Im}}[C_8^{(')NP}]  = 10^{-3}$ \eqref{eq:ref}.
 Uncertainties are in the $50\%$-range, cf. \SEC\ref{sec:directCP}.
 The SM contribution is negligible, down by an
order of magnitude (cf. App.~\ref{app:ACPSM}).
A useful aspect is that the Wilson coefficients of the two chiralities of the chromomagnetic operator 
enter with different sensitivity in \eqref{eq:summary} which has discrimination potential.

The chirality of the photon is an interesting aspect and deserves some discussion in comparing it to 
the $b$-sector.
In $b \to (d,s) \gamma$ transitions the left-handed amplitude dominates 
over the right-handed amplitude as a result of the large $b$-quark mass and 
the $V$-$A$ interactions. 
This pattern might be broken by physics beyond the SM and can be measured in TDCP-asymmetries \cite{Atwood:1997zr}. 
The situation in $D^0 \to V \gamma$ is rather different. 
Whereas it is still true that the left-handed amplitude is larger than the right-handed amplitude, e.g. \eqref{eq:ALD} 
it is not very significant since the $c$-quark mass is smaller.
This neither-nor situation has  consequences. 

Since the amplitudes themselves are LD dominated the TDCP-asymmetries 
are not sensitive to novel right-handed currents.
However, TDCP-asymmetries measure the LD 
chirality asymmetry $\chi_{LD}$ \eqref{eq:chiLD} and thus can provide interesting 
information on LD dynamics and could serve as validation criteria for theoretical  tools. 
Let us add that 
the  feasibility of the measurement depends on the definite value of the mixing phase $\phi_D$ 
(as commented on at the end \SEC\ref{sec:timeCP}).

On the speculative side it is of course possible that NP contributes to SM or non-SM operators
of the WA-type, $O_{1,2}^{d}$ \eqref{eq:Heff}\footnote{Note that in \cite{Isidori:2011qw} 
it is the GIM-combination \eqref{eq:unitarity}, $O_{1,2}^{d} -O_{1,2}^{s}$, 
which is severely constrained through $\epsilon'/\epsilon$ in new weak phases 
but not the individual operators  $O_{1,2}^{d(s)}$ of down and strange per se.}  possibly with new weak phases. 
Allowing for the latter and parametrising a strong phase
for the yet to be computed ${\cal O}(\alpha_s)$-corrections $l_{L(R)} 
\to l_{L(R)} e^{i \Phi_{L(R)}}$, one gets,
\begin{equation}
H[S] =   \chi_{\textrm{LD}} \cdot \left(  - \xi \cos[\sin](\phi_{D} - \Phi_L - \Phi_R) \cos(\delta_L - \delta_R) 
 \right)\;,
 \end{equation}
 and of course  $\chi_{\textrm{LD}}$  is then affected by the NP and needs reevaluation.

At last let us give an outlook and hint how the current work could be improved.
On the experimental side, the  measurement of  the branching ratios of the three charged modes 
 $D^+_d \to  \rho^+ \gamma$ and 
 $D^+_s \to ( \rho,K^{*})^+ \gamma$ would be helpful.
This is the case since  the corresponding Wilson coefficients are not colour suppressed at LO 
and would thus allows us to assess the matrix elements  themselves. 
 On the theoretical side it would benefit from ${\cal O}(\alpha_s)$-correction
 of the WA contributions.  
  In particular the radiative corrections would allow an estimate of the strong phase 
 and the inclusion of the $C_7$-effect \cite{Isidori:2012yx}.
The prominence of  WA  in the isospin asymmetry in $b \to s$ processes provides yet another motivation 
for their reassessment. 
Furthermore
it might  be interesting to extend this work from $D \to V \gamma$ to $D \to V \ell^+\ell^-$ as the latter
might be easier to deal with at the LHCb where the photon final state remains challenging at present. 

In conclusion charm physics is theoretically challenging and the situation with regards to new physics 
remains inconclusive in the sense that it is far from impossible that new physics is lurking in this sector. 
Charm physics in $b \to u$  and non-FCNC modes 
therefore deserves further study in our opinion.
 

\section*{Acknowledgments}

We are grateful to  Ikaros Bigi, Max Hansen, Alex Lenz, Franz Muheim, Steve Playfer, Stefan Recksiegel, Mark Williams, Yuehong Xie and especially Jernej Kamenik, Gino Isidori and Ayan Paul for discussions and/or
 correspondence.
RZ is supported by an STFC Consolidated Grant, ST/P0000630/1.

\appendix

\section{Long-distance Amplitudes}
\label{app:long}

This appendix is devoted to aspects of WA which we argue is the dominant mechanism.

\subsection{Theory: weak annihilation vs quark loops}
\label{app:WAvsQL}

One may distinguish two types of LD contributions according to whether the
 quark level transitions is $ c \bar u \to d \bar d$ or $c \to u  d \bar d(s \bar s)$. They can be  generated by 
 the weak operators ${\cal O}_{1,2}^{d,s}$ \eqref{eq:Heff}  for instance.
 From the viewpoint of quarks and gluons the
 first type  is known as  WA (Fig.~\ref{fig:graphs}, left) and the quark loop (QL) (Fig.~\ref{fig:graphs}, centre;right). 
The WA contributions have been computed  $B,D \to V \gamma$ and $D \to V \gamma$ in \cite{Greub:1996wn,Ali:1995uy,Lyon:2013gba} at ${\cal O}(\alpha_s^0)$.
Note that the  QL  of the type shown in Fig.~\ref{fig:graphs}(centre,right)  are 
evaluated in an $1/m_c(1/m_b)$-expansion for $c(b) \to u(d,s) \gamma$,
although in principle one could compute them in the exclusive case with 
LCSR which is not based on a $1/m_c(1/m_b)$-expansion.

We advocate that  WA  dominates over QL for the following reason.
QL and WA  are generated by the same weak operator, ${\cal O}_{1,2}^{d,s}$ and ${\cal O}_{1,2}^{d}$ \eqref{eq:Heff}  respectively,
yet the QL  is down by two loops with respect to WA.
 This is  the case because the single QL Fig.~\ref{fig:graphs}(centre) vanishes
 by  gauge invariance. The reason therefore is that the photon polarisation $\Pi_{\mu \nu}(q) = (q^2 g_{\mu\nu} - q_\mu q_\nu) \Pi(q^2)$ vanishes for $q^2 =0$ when contracted with the photon polarisation tensor.
Note that in addition  there is a GIM suppression of the QL, 
though not very effective for the matrix elements \cite{Greub:1996wn}.
This suggests a natural hierarchy WA $\gg$ QL in the types of charm 
transitions discussed in 
this paper.\footnote{In the approach in \cite{Burdman:1995te}
the two transitions are modelled with hadronic data. 
We identify  WA and QL with  the pole (P)  and   the vector-meson dominance (VMD) part respectively, 
The comparable numbers for P and VMD are  not in line with  the arguments above.
(We further note that in  \cite{Burdman:1995te} the P-part receives no contribution in ${\cal A}_\parallel ( \leftrightarrow A_{\textrm{PV}})$ which 
is not reflected in the LCSR computation \cite{Khodjamirian:1995uc,Ali:1995uy,Janowski:2021yvz}.) 
A possible issue  is that the signs of the couplings of the VMD models are not known, that is to say 
only their absolute values can be inferred  from experiment.
Thus the formalism might overestimate the contributions as it cannot capture cancellations, 
which gauge invariance suggests to be present.
A similar point of view has been taken in \cite{Burdman:2003rs} 
by one of the authors of \cite{Burdman:1995te} in chapter 3.1.3.}

Some confirmation can be found in  $B$-physics.  That is taking numbers from \cite{Ball:2006eu} for  WA 
and QL  one gets: $|{\cal A}_{QL} / {\cal A}_{WA}|_{B^- \to \rho^- \gamma} \approx 2 \cdot 10^{-2}$.\footnote{WA is Cabibbo suppressed with respect to QL
in $B$-physics. In comparing the WA and QL processes/diagrams we, of course, do not take 
CKM hierarchies into account, especially because they are not present in the charm decays we are interested in.} 
To be more precise, for ${\cal A}_{QL}$ we have taken the charm loop contribution where the 
gluon is radiated into the final state vector-meson.\footnote{Note that WA for $B^0 \to \rho^0 \gamma$ is accidentally small because of cancellations
between tree-level and penguin four quark operator contributions.
We do not expect the same to take place for $D^0 \to (\rho^0,\omega) \gamma$ 
since those cancellations are between tree and penguin four quark operator contributions
and the latter are tiny in $D$-physics.} 

Does this hierarchy remain  intact for  $D$-physics? Despite the obvious fact 
that the $\alpha_s(m_c)$-expansion and the $1/m_c$-expansion are less trustworthy, 
it seems hard to see how a    of two order of magnitude hierarchy  can be overthrown. 
Taking the contribution Fig.~\ref{fig:graphs}(right) for the QL from \cite{Greub:1996wn}, which does 
rely on $1/m_c$-expansion, and the estimates of \cite{Janowski:2021yvz} one gets a number,
$|{\cal A}_{QL} / {\cal A}_{WA}| \approx 2 \cdot 10^{-2}$, which is somewhat accidentally close to the one 
for the $B^- \to \rho^- \gamma$.

\subsection{Experiment: weak annihilation vs quark loops}
\label{app:exWA}

 Let us turn to experiment. 
The known  branching fractions are given by \cite{ParticleDataGroup:2020ssz,Belle:2016mtj}\footnote{
For $D^0 \to \rho^0 \ga$ we have taken the value from Belle \cite{Belle:2016mtj} as this 
is the single measurement and it somewhat unclear to us why \cite{ParticleDataGroup:2020ssz} quotes $1.82(32)\cdot   10^{-5} $.}
\begin{equation}
\label{eq:BF}
{\cal B}(D^0 \to \{ \rho^0,\phi,\bar{K}^{*0} \}  \ga)  =  \{ 1.77(32), 2.81(19), 41(7) \} \cdot   10^{-5} \;,
\end{equation}
with respective uncertainties of $\{17,7,17\}\%$ respectively and the hierarchy is based on the Wolfenstein suppression 
of $\{ \la, \la,\la^0 \}$ at the amplitude level. A crucial feature is that only the $D^0 \to \rho^0 \ga$ amplitude allows for a QL topology.
Hence by comparing the branching fraction rescaled by CKM-factors and wave function decomposition one would  expect  to find  
values compatible with $SU(3)$-flavour symmetry. Let us check and define the auxiliary quantity 
\begin{equation}
x_V \equiv  \frac{c_V^2}{ \la_{CKM} f_V^2 m_V^2} (|{\cal A}_\perp|^2 +|{\cal A}_\parallel|^2) \;,
\end{equation}
where  $c_{\rho^0} = \sqrt{2}$ (and unity otherwise) compensates for the wave function decomposition of the 
$\rho^0 \sim \frac{1}{\sqrt{2}}(\bar u u - \bar dd)$. 
Using  \eqref{eq:BF} one then gets 
\begin{equation}
\frac{1}{x_{\phi }} \{ x_{\rho^0}, x_{\phi}, x_{\bar{K}^{*0}} \} \approx \{ 1.65, 1.00, 0.97 \} \;.
\end{equation}
Now, for $D^0 \to \phi \ga$ and $D^0 \to \bar{K}^{*0}\ga$ the major effect of $SU(3)$ seems to be carried by the decay constants. 
The $D^0 \to \rho^0 \ga$ channel differs and indicates a $28\%$ correction ($1.28^2 \approx 1.65$)
in the amplitude.  We identify the following possible   reasons therefore:
\begin{enumerate}
\item It could be that after all, the doubly loop suppressed QL contribution is sizeable. 
One could argue that the colour suppression of the WA process at LO essentially acts like a loop 
suppression. At  the charm scale a one loop correction could easily amount to $20\%$.
\item The $\rho^0$ is a broad state and it could be that it was not treated uniformly in the experiments 
of $\rho^0 \to e^+ e^-$ from where the decay constant is extracted versus the $D^0 \to \rho^0 \ga$ measurement per se. 
Cf. \cite{Bharucha:2015bzk} where this aspect is stressed in the context of the form factor computation. 
\item There is just a single measurement of this mode and confirmation by another facility 
would be most helpful. Although
 the Belle measurements of the other modes is in line with previous measurements (even though the  $K^*$-node is slightly higher).
\end{enumerate}
We would think that point 1 is the most likely explanation but ultimately we cannot tell. 

\subsection{LCSR vs weak annihilation from experiment}
\label{app:LCSR}

In this section we pose the question whether LCSR can accommodate the $D^0 \to V \ga$ branching fractions
quoted above.  There are two parts to it, the matrix elements and the Wilson coefficients. 
In the neutral case both are problematic at LO due to scale uncertainties. 
Let us discuss them one by one.

\subsubsection{Matrix elements at leading order (in LCSR)}

In the SM at LO in $\al_s$  WA is given by the so-called initial state radiation (cf. Fig.~\ref{fig:graphs} (left)) as the emission from the final state is suppressed by the (light) quark masses.  The former is then simply given by the  $V^{D \to \ga} _{\perp,\parallel}(m_V^2)$ transition form factor. 
This conclusion is also true in QCD factorisation. 
 For the actual form factors  we take the analytic results of a NLO LCSR computation into account 
\cite{Janowski:2021yvz}.\footnote{Note this of course does not mean that WA is covered 
at NLO as it would involve the connection of a gluon between initial and final state quarks which is laborious task.}$^,$\footnote{Of course it is interesting to compare to the earlier computation 
\cite{Khodjamirian:1995uc} which is though LO whereas \cite{Janowski:2021yvz} includes radiative corrections and further higher twist corrections.
The results in $V_\perp$ are comparable to  \cite{Khodjamirian:1995uc} but a little lower. 
There are significant differences for $V_\parallel$ especially in the charged case.
Differences are due to higher twist terms and in the charged case where the difference is largest, ca a factor of $3$, 
this is further due to the non-subtraction of the contact term.} 
The values are collected in the main text  in \eqref{eq:ALD} as they constitute a new result.

\subsubsection{Size of the (effective) Wilson coefficients}

Let us consider the operators 
\begin{equation}
O^0= \bar c\ga_\mu  (1-\ga_5) D \bar D \ga_\mu  (1-\ga_5) u \;,\quad 
O^+ = \bar c\ga_\mu  (1-\ga_5) u \bar D \ga_\mu  (1-\ga_5) D \;,
\end{equation}
(with $D = d,s$) that  govern the weak annihilation transition of $D^{0(+)}$ at LO in $\al_s$. They relate to 
the combination of Wilson coefficients denoted by  
\begin{equation}
H^{\textrm{eff}} \propto a_{2(1)} O^{0(+)} \;, \quad a_2 = C_1/3 + C_2 \;,  \quad a_1 = C_1 + C_2/3 \;,
\end{equation}
and are referred to as colour-suppressed and -allowed respectively. 
The values at the charm scale are  $C_1 \approx 1.2$ and $C_2 \approx 0.4$ e.g. \cite{Brod:2011re} (cf. \cite{Buras:1994ij} for a more elaborate discussion and analysis) 
which lead to $a_2 \approx 0$ and $a_1 \approx  1$.
 Their values at the electroweak scale are of course $C_1 = 1 $ and $C_2 =0$ and $a_2 = 1/3$ 
and $a_1 = 1$ respectively. One concludes that the renormalisation group running for $a_1$ is moderate and can be trusted much to the contrary 
to $a_2$. Its value at the charm scale is absurdly small  (compare  $a_2(m_b) \approx 0.2$).  With such steep running it is clear that the radiative corrections are large. 
The value of $a_2 \approx -  0.5 $ in \cite{Buras:1994ij,Khodjamirian:1995uc},  as used in the main text, is meant to model this effect. Such values were fitted to experiment in other contexts. 

This unsatisfactory situation could sbe improved by  computing the radiative corrections to the WA matrix elements and/or
measuring the charged modes where radiative corrections can be expected to be more moderate. 
However even for the charged modes there is a twist in that for the photon emitted from the charged meson, which is the dominant process at LO, 
there is a large suppression between the charm and strange quark contribution \cite{Janowski:2021yvz}.  
This puts even more pressure on the community to compute WA at next-leading order in $\al_s$.

\subsubsection{Branching fractions and LCSR amplitude}

One may subject the  amplitude \eqref{eq:AX} with Wilson coefficients and form factors as described above 
 to experiment.   
We do so by  comparing  ${\cal B}(D^0 \to  \phi \gamma)$ (suitable as the $\phi$ is narrow) which 
shows that  the   LCSR predictions \cite{Janowski:2021yvz,Khodjamirian:1995uc} 
 differ by about a factor of two. This is not a small effect,  yet not  impossible 
in view of  experimental and theoretical uncertainties. 
 Being pragmatic we  scale 
the neutral modes by the factor $\kappa_0 = 2$ (and $\kappa_+ =1$ in the absence of better knowledge).\footnote{
We note that the  $D^0 \to \rho \ga$ in QCD factorisation  would necessitate $\kappa_0 \approx 3$ when inspecting table I in \cite{deBoer:2017que}. Since the same 
effective Wilson coefficients are assumed in our and their work this means that our LCSR result is ca $50\%$ larger than the QCD factorisation contribution. 
This is well within the expected ballpark since already for $b$-physics the correction in $1/m_b$ are sizeable cf. Sec. 5 \cite{Dimou:2012un}.}  Such procedures are not ideal but there is no other way at present.


\subsection{The $D_{(s)}^+ \to V \ga$ branching fractions}
\label{app:overview}

Here we give an overview of the main $D \to V \ga$ modes collected in  Tab.~\ref{tab:decays}. The neutral ones are all measured but not the 
charged ones.
\begin{table}[h]
\begin{center}
\begin{tabular}{c |  l | l |  l |  l  |  c  | c | r }
No. & decay & FCNC& transition & $O_8$ & CKM & cs & ${\cal B}(D \to  V \gamma)$  \\[0.1cm] \hline
1 & $D^0 \to \rho^0(\omega) \gamma$ & $c \to u$ & $c\bar u  \to d \bar d$ & yes & $\lambda^1$  & yes  & $1.77(32) \cdot10^{-5}$  \cite{Belle:2016mtj} \\[0.1cm]
2 &$D^0 \to \phi  \gamma$ &  $c \to u$ & $c \bar u  \to s \bar s$ & no & $\lambda^1$  &  yes &$2.81(19) \cdot10^{-5}$  \cite{ParticleDataGroup:2020ssz} \\[0.1cm] 
3 &$D^0  \to \bar K^{0,*}  \gamma$ &    no & $c \bar u   \to s \bar d$ & no & $\lambda^0$  &  yes & $4.1(7) \cdot10^{-4}$   \cite{ParticleDataGroup:2020ssz}\\[0.1cm]  \hline \hline
4 &$D^+ \to \rho^+ \gamma$ & $c \to u$ & $c\bar d  \to u\bar d$ & yes & $\lambda^1$  &  no & $6.4 \cdot 10^{-6}$ this work  \\[0.1cm] 
5 &$D_{s}^+ \to K^{*+} \gamma$ & $c \to u$ & $c \bar s  \to u \bar s$ & yes & $\lambda^1$  &  no & $1.7  \cdot10^{-5}$  this work \\[0.1cm] 
6 & $D_{s}^+ \to \rho^+ \gamma$ &  no & $c \bar s  \to  u \bar d $ & no & $\lambda^0$  &  no & $2.1 \cdot 10^{-4} $   this work
\end{tabular}
\end{center}
\caption{\small The Wolfenstein parameter is $\lambda \approx 0.23 $ and the acronym cs stands for  colour suppressed.
  Decays 3 and 6 are not of the FCNC type, in the sense that they can be directly written in terms 
of tree $W$-exchange. The $O_8$ column indicates whether they have $O_8$ matrix elements and are thus potentially 
 CP-violating in the context of this paper.  
The experimental value in the first row is for the $\rho^0$-case as for the $\omega$ only a bound exists at present  \cite{ParticleDataGroup:2020ssz}.
Additionally, we refer the reader to table I and II in \cite{deBoer:2017que} where a comparison between different studies for the branching fractions has been made which includes QCD factorisation applied to charm decays.}
\label{tab:decays}
\end{table}
Here we give reference values using the  values in \eqref{eq:ALD}, taken from  \cite{Janowski:2021yvz} 
(and  use $\kappa_+ =1$ in  \eqref{eq:AX} and  $a_1 = 1$ as by  above).
With this input we get the values 
quoted in Tab.~\ref{tab:decays}. The hierarchies   are easily understood in our approximation where the basic amplitude is degenerate 
\begin{equation}
{\cal B}(D^+ \to \rho^+  \ga) = \frac{\tau_{D^+}}{\tau_{D_s}} {\cal B}(D_s \to K^{*+}  \ga)  =  \la^2 \frac{\tau_{D^+}}{\tau_{D_s}} {\cal B}(D_s \to \rho^{+} \ga) \;.
\end{equation} 
 Above $\la  \approx 0.266 $  is the Wolfenstein parameter and the ratio of lifetimes is $\frac{\tau_{D^+}}{\tau_{D_s}}  \approx 2$ because of the pion's wave function decomposition.  The uncertainty at the amplitude level is easily $50\%$ (improvable  by an NLO computation  as 
already mentioned a few times).

\section{Formulae for  CP-violation}
\label{app:CPbasic}

\subsection{Formulae for direct CP-violation}

In this appendix we collect some formulae which are useful throughout the text.
We shall parametrise  an amplitude as follows, 
\begin{equation}
\label{eq:dec}
{\cal A}(D^0 \to f) = A_a e^{i\delta_a} e^{i\phi_a} 
+ A_{b} e^{i\delta_b}e^{i\phi_b} \quad ,
\end{equation}
with weak (CP-odd) phases $\phi$ and  strong (CP-even) phases $\delta$ separated
to leave $A_{a,b}$ real. 
Note that in the SM the decomposition \eqref{eq:dec}  is sufficient as one might use unitarity \eqref{eq:unitarity} to eliminate one amplitude  to arrive at two amplitudes.
Using the notation $\Delta \equiv \frac{A_a}{A_b}$ , $\delta(\phi)_{ab} = \delta(\phi)_a-\delta(\phi)_b$ the CP-asymmetry becomes:
\begin{eqnarray}
\label{eq:ACP}
A_{CP}[ D^0 \to f] = \frac{- 2  \sin(\delta_{ab}) \sin (\phi_{ab}) \Delta }{1+ 2 \Delta \cos(\delta_{ab}) \cos(\phi_{ab}) + \Delta^2 }  \stackrel{\Delta \ll 1}{\approx} \!\!  - 2  \sin(\delta_{ab}) \sin (\phi_{ab}) \Delta \;.
\end{eqnarray}
In the second line we have assumed a hierarchy between the amplitudes which is the case 
for $D^0 \to (\rho^0,\omega) \gamma$ as studied in this paper.

\subsection{Formulae for TDCP-violation}
\label{app:time}

The replacement due to the relevance of ${\cal O}_7$ as described in subsection 
\ref{sec:timeCP} is as follows:
\begin{eqnarray} 
\label{eq:tg}
g_L  e^{i\delta} e^{i \phi_L } \to    \tilde g_L e^{i \Delta_L}   e^{i \Phi_L} 
  =  \frac{G_F}{\sqrt{2}} \left( \frac{e m_c }{2 \pi^2} \right)   \frac{1}{c_V}  [C_8  (2 G_1(0)) + C_7   (2 T_1(0))]  \;,
\end{eqnarray}
and for $g_R$ is given by the following replacements: $L \to R$ and $C_8,C_7 \to  C_8',C_7'$.
Note that unlike before we cannot assume a common strong phase as the ratios $C_8/C_7$
and $C_8'/C_7'$ might not necessarily be the same. This is why the strong phase $\Delta$ carries
a chirality label. The symbol $\Phi$ denotes the weak phase.
The formulae for $H$ and $S$ in \eqref{eq:CPform} are given, including a derivation, 
 in the appendix of reference \cite{Muheim:2008vu}\footnote{Note that the different sign of $H$ as wrt \cite{Muheim:2008vu}. This originates from the fact that $\Delta \Gamma_s$ in that reference is the light minus the heavy decay rate rather
 than the other way around as is assumed in the $D^0$-system. The reason for this difference in convention 
 is that in each case $\Delta \Gamma$ is chosen to be positive. Of course  this sign 
 is experimentally unobservable as only $\Delta \Gamma \times H^{2n+1}$ for integer $n$ is observable.} and take the following form:
\begin{eqnarray}
 H[S]  &\;=\;&   \frac{ - 4 \xi }{n} \Big( l_L l_R  \cos(  \delta_L - \delta_R)\cos[\sin](\phi_{D})     \nonumber \\[0.1cm]
&\;+\; &  \tilde g_L \tilde g_R \cos(\Delta_L \mi \Delta_R )
\cos[\sin]( \phi_{D} \mi \Phi_L \mi \Phi_R)   \nonumber \\[0.1cm]
&\;+\; &  (  \tilde g_L l_R \cos(\Delta_L - \delta_R) \cos[\sin]( \phi_{D} \mi \Phi_L)  +  \{ L \leftrightarrow R \}  )  \Big)  \;, \nonumber 
\end{eqnarray}
with 
$$
 n  \equiv 2( l_L^2 + 2  \left( g_L l_L \cos(\Delta_L- \delta_L)  \cos(\phi_L)   g_L^2 \right) +   \{ L \leftrightarrow R \}  )  \;,
$$
and where $\xi$ is the CP-eigenvalue of $V$. 
For $V = \{\rho,\omega,\phi,\bar{K}^*(\bar{K}_S\pi^0)\}$ the eigenvalue is $\xi  = 1$ and
for $V=\bar{K}^*(\bar{K}_L\pi^0)$ it is $\xi  = -1$.

\section{ $A_{CP}( D^0 \to V \gamma )$ other than through $C_8^{\textrm{NP}}$ }
\label{app:comment}

For our discussion it is convenient to write the amplitude as follows,
\begin{equation}
{\cal A} \approx \lambda_d e^{i \delta _d } A_d  +  \lambda_s e^{i \delta _s } A_s  +  \lambda_b e^{i \delta _b } A_b  \;,
\end{equation}
which is similar to \eqref{eq:dec} with the exception that the unitarity relation \eqref{eq:unitarity}
has not been used and that the weak phases are contained within $\lambda_{d,s,b}$ \eqref{eq:Heff_generic}.
As argued in App.~\ref{app:long} we expect  the lion's share of  $A_d$
to be covered by WA which has, presumably, a small strong phase which we shall neglect ($\delta_d \to 0$). 
We assume a Wolfenstein parametrisation up to order ${\cal O}(\lambda^5)$ which
fulfils, e.g. \cite{Donoghue:1992dd},
 \begin{equation}
 \label{eq:Image}
 {\textrm{Im}}[ \lambda_d ] = 0 \:, \quad  {\textrm{Im}}[ \lambda_s ]  = A^2\lambda^5\eta \;, \quad  {\textrm{Im}}[ \lambda_b ] = - A^2 \lambda^5 \eta
 \end{equation}
where $A$, $\rho$ and $\eta$ are the other three Wolfenstein parameters and $A^2 \lambda^5 \eta \approx
1.4 \cdot 10^{-4}$. Eq.\eqref{eq:Image}.
The fact that $|{\textrm{Im}} [\lambda_{b,s}] | \approx 1.4 \cdot 10^{-4}$  indicates small CP-asymmetries\footnote{One might be tempted to say that if  WA  dominates by another two order of magnitudes  then
this implies that the CP asymmetry is automatically below $10^{-5}$. This is not correct as in this way of thinking the absolute value of $\lambda_b$ should be factored into $A_b$ and then ${\textrm{Im}}[\lambda_b/|\lambda_b |] \approx {\cal O}(1)$ is not small any more.}, of that order.

Thus it remains to identify 
contributions with sizeable strong phases $\delta_{s,b}$ and amplitudes $A_{s,b}$ for which we see two major sources.
First the matrix element of ${\cal O}_8$ e.g. \eqref{eq:ImG1} \cite{Dimou:2012un} 
and second the matrix element of ${\cal O}^{d,s}_2$ \cite{Greub:1996wn} (cf. Fig.~\ref{fig:graphs}(right) for a contribution)
giving rise, effectively, to an ${\cal O}_7$-operator. The latter 
as well as its matrix element analogous to \eqref{eq:MO8} are defined and parameterised respectively as follows,
\begin{eqnarray}
\label{eq:O7}
 {\cal O}_7^{(')}  &\equiv& - \frac{ m_c  e  }{8 \pi^2} \,  \bar u \sigma_{\mu\nu}  F^{\mu\nu} (1\pm \gamma_5) c  
 \\[0.1cm]
\label{eq:MO7} 
\matel{V\gamma}{{\cal O}^{(')}_7}{D} &=& \left( \frac{e  m_c }{4 \pi^2} \right) \frac{1}{c_V} (T_1(0) P_\perp \pm T_2(0) P_\parallel) \;.
\end{eqnarray}

\subsection{Effective Wilson coefficients $C_{7,8}^{\textrm{eff}}(m_c)$}
\label{app:Ceff}

Let us state that we do not intend to give a critical review of the treatment of Wilson coefficients 
in the charm sector, e.g. of whether it makes sense to include 
light-quarks into SD contributions evaluated in perturbation theory\footnote{We are grateful to Ikaros Bigi 
and Ayan Paul to draw our attention to this point.}.
We shall simply follow the literature. 
It is fortunate that the SD contributions turn out to be subdominant in the SM.

The different contributions discussed above are conveniently discussed in terms 
of so-called effective Wilson coefficients. The latter 
 consists of the pure Wilson coefficient $C_{7,8}(m_c)$ and matrix elements which 
can be rewritten in terms of ${\cal O}_{7,8}$ which we denote by $\delta C_{7,8}^{\textrm{eff}}(m_c)$
\begin{eqnarray}
C_{7,8}^{\textrm{eff}}(m_c) = C_{7,8}(m_c)  +  \delta C_{7,8}^{\textrm{eff}}(m_c)  \;.
\end{eqnarray}
From a conceptual point of view the
 Wilson coefficient can be divided into two further sub-parts,
\begin{equation}
C_{7,8}(m_c)   = C_{7,8}^{(m_W)}(m_c) +  C_{7,8}^{(m_b)}(m_c) \;.
\end{equation}
The notation above is non-standard but hopefully useful for clarity.
For the reminder of this section we closely follow the notation of \cite{Paul:2011ar}.
For $C_8^{\textrm{eff}}$ only  $C_{8}^{(m_W)}(m_c)  =  \eta_c^{\frac{14}{25}}\eta_b^{\frac{14}{23}}C_8(m_W) $, 
$\eta_b = \alpha_s(m_W)/\alpha_s(m_b)$ and  $\eta_c= \alpha_s(m_b)/ \alpha_s(m_c)$,  is known explicitly in the literature. 
For  $C_7^{\textrm{eff}}$ all three parts are known which we shall quote, almost explicitly, below,
\begin{eqnarray}
\label{eq:C7mc}
C_7^{(m_W)}(m_c)  &=&   \left[  \eta_c^{\frac{16}{25}}\eta_b^{\frac{16}{23}}C_7(m_W) \!-\! \frac{16}{3}\left(\eta_c^{\frac{14}{25}}\eta_b^{\frac{14}{23}} \!-\! \eta_c^{\frac{16}{25}}\eta_b^{\frac{16}{23}}\right)C_8(m_W) \right]  \nonumber \\[0.1cm]
C_7^{(m_b)}(m_c)  &=& -    \lambda_b \sum _{i,j} C_j(m_b)X_{ji} \eta_c^{z_i} \;,
 \end{eqnarray}
 where $i = 1..8$, $j = 1..6$. 
 Note that $C_7^{(m_W)}(m_c)$ describes the evolution directly from 
 $m_W$ to $m_c$ and  $C_7^{(m_b)}(m_c)$ originates from 
integrating out the $b$-quark at the $m_b$-scale and running from $m_b$ to $m_c$.
We hasten to add that the above expressions are given in the leading logarithm approximation.
The term from the four quark matrix element is given by  \cite{Greub:1996wn}
\begin{equation}
\label{eq:deltaC7eff}
\delta  C_{7}^{\textrm{eff}}(m_c) = \frac{\alpha_s(m_c)}{4\pi }C_2(m_c)\left(  \lambda_s f[ (m_s/m_c)^2]+ 
 \lambda_d f[ (m_d/m_c)^2]  \right) \;.
\end{equation}
The strong phase results from the charmed meson's four momentum cutting the diagram through light quark lines.
The contribution of $C_1(m_c)$ vanishes whereas the $C_{3,4,5,6}(m_c)$ have not been given 
but are small as they originate from SD contributions which themselves are small. 
In fact the numerical hierarchy is as follows \cite{Greub:1996wn}:
\begin{equation}
\label{eq:hierarchy}
|C_7^{(m_W)}(m_c)| \approx 2 \cdot 10^{-7} \ll  |C_7^{(m_b)}(m_c)| \approx 8 \cdot 10^{-6} \ll  |\delta C_7^{\textrm{eff}}(m_c)| = 5 \cdot 10^{-3}  \;.
\end{equation}
The hierarchy between the first two was noted in \cite{Burdman:1995te} and numerically improved in \cite{Greub:1996wn}.
The fact that matrix element dominates the Wilson coefficient was pointed out in  \cite{Greub:1996wn}.
The expression of $C_7^{(m_b)}(m_c)$ for operators other than ${\cal O}_{1,2}$ was given recently in ref.\cite{Paul:2011ar}.
As mentioned previously we are not aware of explicit results for
 $C_8^{(m_b)}(m_c)$ and $\delta C_8^{\textrm{eff}}(m_c)$ in the literature,
  yet they can be expected to be close 
 to their $C_7$-counterparts as they differ only by colour factors.
 Excluding cancellation 
 effects we would  expect them to equal up to ${\cal O}(1/N_c)$ effects, say equal to 
 about $30-50\%$. Given the uncertainties of the estimates 
 the approximations, $C_8^{(m_b)}(m_c) \approx   
 C_7^{(m_b)}(m_c)$ and $\delta C_8^{\textrm{eff}}(m_c)  \approx   
 \delta  C_7^{\textrm{eff}}(m_c) $,   are 
 good for our purposes\footnote{Though the values $C_{7,8}^{(m_W)}(m_c)$ differ substantially for various reasons but 
this is of no concern as they are small.}.  
Furthermore with $C_8(m_c) \approx C_8^{(m_b)}(m_c) \approx C_7^{(m_b)}(m_c)  \approx (-0.3+ 0.8i)  \cdot 10^{-5}$ 
we see that the SM value is two to three order of magnitude below the reference value $ {\textrm{Im}}[C_8^{NP}] \approx 
0.4 \cdot 10^{-2}$.

 \subsection{$A_{CP}( D^0 \to V \gamma )$ in the SM}
 \label{app:ACPSM}
 
 In the SM we identify three main sources contributing to the direct CP-asymmetry: a) $C_8(m_c) \approx 
 C_8^{(m_b)}(m_c)$ b) $\delta C_7^{\textrm{eff}}(m_c)$ and c) $\delta C_8^{\textrm{eff}}(m_c)$. 
 Right-handed operators   
${\cal O}_{7,8}^{(')}$ are negligible in the SM
as Wilson coefficients as well as matrix elements are 
suppressed.
  As previously mentioned we shall use $C_8(m_c) \approx C_7(m_c)$  for  
 cases a) and c) which is good up to $1/N_c$ corrections.
 Note, as the leading LD amplitude is proportional to $\lambda_d$, it is only $\lambda_s$ or $\lambda_b$ that 
 can contribute to the direct CP-asymmetry.  
 \begin{itemize}
 \item[a)] It is found that \cite{Greub:1996wn}
 \begin{equation} 
 C_7^{(m_b)}(m_c)  \approx   0.06  \lambda_b  \approx (0.3 - 0.8 i) \cdot 10^{-5}
 \end{equation}
 and assuming, as discussed above,   $C_8^{(m_b)}(m_c) \approx C_7^{(m_b)}(m_c)$, we get that this contribution 
 compares with $C_8^{\textrm{NP}}$ in $A_{\textrm{CP}}$ as follows:
 \begin{equation}
 \frac{0.06 \, {\textrm{Im}}[  \lambda_b]}{ {\textrm{Im}}[C^{\textrm{NP}}_8]} \approx  - 0.2 \cdot 10^{-2} \;.
 \end{equation}
 
 \item[b)]
 It is found that
 \begin{equation}
 \label{eq:C7effN}
 \delta C_7^{\textrm{eff}}(m_c) =  (0.6  + 2.2 i)\cdot 10^{-2} \,  \lambda_s  + c \lambda_d  \;,
 \end{equation} 
where the imaginary part, other than $ \lambda_s$, corresponds to a strong phase.  
The number $c$ is of no importance for CP-violation as it can be absorbed into WA which is 
proportional to $\lambda_d$ and much larger.
The contribution $A_{\textrm{CP}}$ compares with 
$C_8^{\textrm{NP}}$ as follows:
\begin{equation}
\label{eq:2}
\frac{ {\textrm{Im}}[\lambda_s ]  {\textrm{Im}}[ (0.6  + 2.2 i)\cdot 10^{-2} ] T_1(0)}{{\textrm{Im}}[C^{\textrm{NP}}_8]   {\textrm{Im}}[G_1(0)]}  \approx - 1 \cdot 10^{-2} \;,
\end{equation}
for reference values \eqref{eq:NC8},  $T_1(0) = 0.7$ and $ {\textrm{Im}}[G_1(0)]  = -0.2$.
\item[c)] As discussed above we expect $\delta C_8^{\textrm{eff}}(m_c) \approx  \delta C_7^{\textrm{eff}}(m_c)$ 
and this leads to a result for c) with  ${\textrm{Im}}[G_1(0)]/T_1(0)  \approx 2/7$ suppression factor as compared to 
\eqref{eq:2}.
\end{itemize}
Summa summarum the SM contributions is one order of magnitude below the values
${\textrm{Im}}[C^{\textrm{NP}}_8] $ \eqref{eq:NC8}-contribution and with the value in \eqref{eq:ACPexplicit} we get
\begin{equation}
\label{eq:ACPSP}
A_{\textrm{CP}}|_{\textrm{SM}}(D^0 \to (\rho^0,\omega) \gamma  ) \approx   (-1.5 \% \frac{1}{\sqrt{3}}) (  - 2 \cdot 10^{-2}) \approx 3  \frac{1}{\sqrt{3}} \cdot 10^{-4}  \;.
\end{equation}
We refrain from quoting a specific uncertainty. We would though think that the value catches the right order 
of magnitude.  As possible criticisms one could advocate for example the estimate $C_8^{(m_b)}(m_c) \approx C_7^{(m_b)}(m_c)$ and question 
the accuracy of local duality in \eqref{eq:C7effN}.
 The charged case is obtained by replacing ${\textrm{Im}}[G_1^{D^0}] \to  {\textrm{Im}}[G_1^{D^+}]$ in \eqref{eq:2} and this would lead to $ A_{\textrm{CP}}|_{\textrm{SM}}^{D^+} \approx 3.9\% \frac{1}{\sqrt{3}} 
(-3 \cdot 10^{-2}) \approx - 1 \frac{1}{\sqrt{3}} \cdot 10^{-3} $.

 \subsection{$A_{CP}( D^0 \to V \gamma )$  via ${\textrm{Im}}[ C_7^{\textrm{NP}}]$ and a strong LD-phase}
 \label{app:ACP87}

In reference \cite{Isidori:2012yx} the idea was put forward  
that $C_8(m_{\textrm{NP}})$ mixes into $C_7(m_c)$, e.g. Eq.~\eqref{eq:C7mc} for the SM evolution. 
More precisely depending on the model and the scale of NP, $M_{\textrm{NP}}$,  it was put forward  \cite{Isidori:2012yx} that this  leads to  comparable values\footnote{Note that our normalisation of ${\cal O}_7$ differs from \cite{Isidori:2012yx} by a factor of $Q_u$ which translates 
in to $Q_u C_7 = C_7^{IK}$, where $IK$ stands for the authors of \cite{Isidori:2012yx}.}.  An important point 
is that $C_7(m_c)$ hardly affects $D^0 \to \pi\pi/KK$ because of $\alpha$-suppression 
and is therefore not constrained by the latter. 
Following \cite{Isidori:2012yx} we shall assume only SM degrees of freedom below  the scale $M_{\textrm{NP}} = 1 
\TeV$ and that the NP part of the Wilson coefficients is much larger than the SM part. 
 Amending the notation  of \eqref{eq:C7mc} to include the running of six quarks above the top threshold one gets 
 \begin{eqnarray}
  \label{eq:rel}
 C_8^{(1 \TeV)}(m_c) &\approx&  0.42 C_8(1 \TeV)  \;,  \nonumber  \\[0.1cm]
  C_7^{(1 \TeV)}(m_c) &\approx&  0.37 C_7(1 \TeV) - 0.26 C_8(1 \TeV) 
  \nonumber  \\[0.1cm] 
  &\approx&  0.37 C_7{(1 \TeV)} - 0.62 C_8(m_c) \;,   \nonumber
  \end{eqnarray}
and the analogous equations for the ${\cal O}_{7,8}^{'}$-operators. Eq.~\eqref{eq:rel} exposes the dependence of $C_7(m_c)$ on  the  scale $M_{\textrm{NP}}$ and $C_7^{(')}(M_{\textrm{NP}})$. 
We shall somewhat arbitrarily choose the value $ {\textrm{Im}}[C_7^{\textrm{(')NP}}(m_c)]  \approx  -0.5 {\textrm{Im}} [C_8^{\textrm{(')NP}}(m_c)]$ as a reference values.
This follows  the model dependent assumption $ |{\textrm{Im}}[C_7^{(')}(1 \TeV)]| \ll 
 |{\textrm{Im}}[C_8^{(')}(1 \TeV)]|$ 
 in \cite{Isidori:2012yx}.

Since  the ${\cal O}_7$ matrix element itself, as opposed to $\delta C_7^{\textrm{eff}}$,
does not carry a strong phase and the LD strong phase vanishes at leading order in 
the chiral limit, as discussed in \APP\ref{app:LCSR}, we did not include this effect in our results (\ref{eq:ACPexplicit},\ref{eq:ACPexplicit2}).  In fact we estimated that the phases could be around
 $|\delta_{L,R}| \approx 10^\circ$ and we shall investigate how the CP-asymmetry changes.
It is then useful to rewrite the $g_L$ amplitude as in \eqref{eq:tg} with the replacement:
\begin{equation}
 [C_8  (2 G_1(0)) + C_7   (2 T_1(0))]  \to   [2  {\textrm{Im}}[C_8]  (\underbrace{ G_1(0) -  0.5    T_1(0)}_{F_1})]
\end{equation}
For $T_1(0) = 0.7$ and $G_1^{D^0}(0) \approx -0.2-0.2 i  \approx 0.3 e^{- i  135^\circ}  $ \cite{Dimou:2012un} one gets 
$F_1 \approx  - 0.55 - 0. 2 i = 0.7 e^{- i  160^\circ}$. Thus a  correction of the LD phase  
$\delta_{L,R} = \pm 10^\circ$  leads to a strong phase difference between the two amplitudes in the range of $10^\circ$ to $30^\circ$  which corresponds to a rescaling
of the CP asymmetry by factors $\sin(10^\circ)/\sin(20^\circ) \approx   0.5$ and $\sin(30^\circ)/\sin(20^\circ) \approx   1.5$ respectively.  Thus in conclusion one cannot 
exclude the possibility that the phases conspire to cancel a significant part, or even an order of magnitude,  of the effect!  
A  lot of things have to go wrong for this to happen though. 
As discussed in \SEC\ref{app:LCSR} an
${\cal O}(\alpha_s)$ computation would presumably give an indication of the sign of the LD phase
as well as its size and would allow to make firmer statements.

\bibliographystyle{utphys}
\bibliography{charmCP.bib}

\providecommand{\href}[2]{#2}\begingroup\raggedright\begin{thebibliography}{10}

\bibitem{LHCb:2011osy}
{\bfseries LHCb} Collaboration, R.~Aaij {\em et~al.}, ``{Evidence for CP
  violation in time-integrated $D^0 \to h^-h^+$ decay rates},''
  \href{http://dx.doi.org/10.1103/PhysRevLett.108.111602}{{\em Phys. Rev.
  Lett.} {\bfseries 108} (2012) 111602},
  \href{http://arxiv.org/abs/1112.0938}{{\ttfamily arXiv:1112.0938 [hep-ex]}}.

\bibitem{CDF:2011ejf}
{\bfseries CDF} Collaboration, T.~Aaltonen {\em et~al.}, ``{Measurement of
  CP--violating asymmetries in $D^0\to\pi^+\pi^-$ and $D^0\to K^+K^-$ decays at
  CDF},'' \href{http://dx.doi.org/10.1103/PhysRevD.85.012009}{{\em Phys. Rev.
  D} {\bfseries 85} (2012) 012009},
  \href{http://arxiv.org/abs/1111.5023}{{\ttfamily arXiv:1111.5023 [hep-ex]}}.

\bibitem{LHCb:2019hro}
{\bfseries LHCb} Collaboration, R.~Aaij {\em et~al.}, ``{Observation of CP
  Violation in Charm Decays},''
  \href{http://dx.doi.org/10.1103/PhysRevLett.122.211803}{{\em Phys. Rev.
  Lett.} {\bfseries 122} no.~21, (2019) 211803},
  \href{http://arxiv.org/abs/1903.08726}{{\ttfamily arXiv:1903.08726
  [hep-ex]}}.

\bibitem{HFLAV:2019otj}
{\bfseries HFLAV} Collaboration, Y.~S. Amhis {\em et~al.}, ``{Averages of
  b-hadron, c-hadron, and $\tau $-lepton properties as of 2018},''
  \href{http://dx.doi.org/10.1140/epjc/s10052-020-8156-7}{{\em Eur. Phys. J. C}
  {\bfseries 81} no.~3, (2021) 226},
  \href{http://arxiv.org/abs/1909.12524}{{\ttfamily arXiv:1909.12524
  [hep-ex]}}.

\bibitem{Hansen:2012tf}
M.~T. Hansen and S.~R. Sharpe, ``{Multiple-channel generalization of
  Lellouch-Luscher formula},''
  \href{http://dx.doi.org/10.1103/PhysRevD.86.016007}{{\em Phys. Rev. D}
  {\bfseries 86} (2012) 016007},
  \href{http://arxiv.org/abs/1204.0826}{{\ttfamily arXiv:1204.0826 [hep-lat]}}.

\bibitem{Bigi:2011re}
I.~I. Bigi, A.~Paul, and S.~Recksiegel, ``{Conclusions from CDF Results on CP
  Violation in $D^0 \to \pi^+\pi^-, K^+K^-$ and Future Tasks},''
  \href{http://dx.doi.org/10.1007/JHEP06(2011)089}{{\em JHEP} {\bfseries 06}
  (2011) 089}, \href{http://arxiv.org/abs/1103.5785}{{\ttfamily arXiv:1103.5785
  [hep-ph]}}.

\bibitem{Rozanov:2011gj}
A.~N. Rozanov and M.~I. Vysotsky, ``{CP violation in D-meson decays and the
  fourth generation},'' \href{http://arxiv.org/abs/1111.6949}{{\ttfamily
  arXiv:1111.6949 [hep-ph]}}.

\bibitem{Cheng:2012wr}
H.-Y. Cheng and C.-W. Chiang, ``{Direct CP violation in two-body hadronic
  charmed meson decays},''
  \href{http://dx.doi.org/10.1103/PhysRevD.85.034036}{{\em Phys. Rev. D}
  {\bfseries 85} (2012) 034036},
  \href{http://arxiv.org/abs/1201.0785}{{\ttfamily arXiv:1201.0785 [hep-ph]}}.
  [Erratum: Phys.Rev.D 85, 079903 (2012)].

\bibitem{Li:2012cfa}
H.-n. Li, C.-D. Lu, and F.-S. Yu, ``{Branching ratios and direct CP asymmetries
  in $D\to PP$ decays},''
  \href{http://dx.doi.org/10.1103/PhysRevD.86.036012}{{\em Phys. Rev. D}
  {\bfseries 86} (2012) 036012},
  \href{http://arxiv.org/abs/1203.3120}{{\ttfamily arXiv:1203.3120 [hep-ph]}}.

\bibitem{Brod:2011re}
J.~Brod, A.~L. Kagan, and J.~Zupan, ``{Size of direct CP violation in singly
  Cabibbo-suppressed D decays},''
  \href{http://dx.doi.org/10.1103/PhysRevD.86.014023}{{\em Phys. Rev. D}
  {\bfseries 86} (2012) 014023},
  \href{http://arxiv.org/abs/1111.5000}{{\ttfamily arXiv:1111.5000 [hep-ph]}}.

\bibitem{Bhattacharya:2012ah}
B.~Bhattacharya, M.~Gronau, and J.~L. Rosner, ``{CP asymmetries in
  singly-Cabibbo-suppressed $D$ decays to two pseudoscalar mesons},''
  \href{http://dx.doi.org/10.1103/PhysRevD.85.054014}{{\em Phys. Rev. D}
  {\bfseries 85} (2012) 054014},
  \href{http://arxiv.org/abs/1201.2351}{{\ttfamily arXiv:1201.2351 [hep-ph]}}.

\bibitem{Feldmann:2012js}
T.~Feldmann, S.~Nandi, and A.~Soni, ``{Repercussions of Flavour Symmetry
  Breaking on CP Violation in D-Meson Decays},''
  \href{http://dx.doi.org/10.1007/JHEP06(2012)007}{{\em JHEP} {\bfseries 06}
  (2012) 007}, \href{http://arxiv.org/abs/1202.3795}{{\ttfamily arXiv:1202.3795
  [hep-ph]}}.

\bibitem{Franco:2012ck}
E.~Franco, S.~Mishima, and L.~Silvestrini, ``{The Standard Model confronts CP
  violation in $D^0 \to \pi^+\pi^-$ and $D^0 \to K^+K^-$},''
  \href{http://dx.doi.org/10.1007/JHEP05(2012)140}{{\em JHEP} {\bfseries 05}
  (2012) 140}, \href{http://arxiv.org/abs/1203.3131}{{\ttfamily arXiv:1203.3131
  [hep-ph]}}.

\bibitem{Golden:1989qx}
M.~Golden and B.~Grinstein, ``{Enhanced CP Violations in Hadronic Charm
  Decays},'' \href{http://dx.doi.org/10.1016/0370-2693(89)90353-5}{{\em Phys.
  Lett. B} {\bfseries 222} (1989) 501--506}.

\bibitem{Isidori:2011qw}
G.~Isidori, J.~F. Kamenik, Z.~Ligeti, and G.~Perez, ``{Implications of the LHCb
  Evidence for Charm CP Violation},''
  \href{http://dx.doi.org/10.1016/j.physletb.2012.03.046}{{\em Phys. Lett. B}
  {\bfseries 711} (2012) 46--51},
  \href{http://arxiv.org/abs/1111.4987}{{\ttfamily arXiv:1111.4987 [hep-ph]}}.

\bibitem{Giudice:2012qq}
G.~F. Giudice, G.~Isidori, and P.~Paradisi, ``{Direct CP violation in charm and
  flavor mixing beyond the SM},''
  \href{http://dx.doi.org/10.1007/JHEP04(2012)060}{{\em JHEP} {\bfseries 04}
  (2012) 060}, \href{http://arxiv.org/abs/1201.6204}{{\ttfamily arXiv:1201.6204
  [hep-ph]}}.

\bibitem{Isidori:2012yx}
G.~Isidori and J.~F. Kamenik, ``{Shedding light on CP violation in the charm
  system via D to V gamma decays},''
  \href{http://dx.doi.org/10.1103/PhysRevLett.109.171801}{{\em Phys. Rev.
  Lett.} {\bfseries 109} (2012) 171801},
  \href{http://arxiv.org/abs/1205.3164}{{\ttfamily arXiv:1205.3164 [hep-ph]}}.

\bibitem{Grossman:2012eb}
Y.~Grossman, A.~L. Kagan, and J.~Zupan, ``{Testing for new physics in singly
  Cabibbo suppressed D decays},''
  \href{http://dx.doi.org/10.1103/PhysRevD.85.114036}{{\em Phys. Rev. D}
  {\bfseries 85} (2012) 114036},
  \href{http://arxiv.org/abs/1204.3557}{{\ttfamily arXiv:1204.3557 [hep-ph]}}.

\bibitem{Grossman:2006jg}
Y.~Grossman, A.~L. Kagan, and Y.~Nir, ``{New physics and CP violation in singly
  Cabibbo suppressed D decays},''
  \href{http://dx.doi.org/10.1103/PhysRevD.75.036008}{{\em Phys. Rev. D}
  {\bfseries 75} (2007) 036008},
  \href{http://arxiv.org/abs/hep-ph/0609178}{{\ttfamily arXiv:hep-ph/0609178}}.

\bibitem{Hiller:2012wf}
G.~Hiller, Y.~Hochberg, and Y.~Nir, ``{Supersymmetric $\Delta A_{CP}$},''
  \href{http://dx.doi.org/10.1103/PhysRevD.85.116008}{{\em Phys. Rev. D}
  {\bfseries 85} (2012) 116008},
  \href{http://arxiv.org/abs/1204.1046}{{\ttfamily arXiv:1204.1046 [hep-ph]}}.

\bibitem{deBoer:2017que}
S.~de~Boer and G.~Hiller, ``{Rare radiative charm decays within the standard
  model and beyond},'' \href{http://dx.doi.org/10.1007/JHEP08(2017)091}{{\em
  JHEP} {\bfseries 08} (2017) 091},
  \href{http://arxiv.org/abs/1701.06392}{{\ttfamily arXiv:1701.06392
  [hep-ph]}}.

\bibitem{Delaunay:2012cz}
C.~Delaunay, J.~F. Kamenik, G.~Perez, and L.~Randall, ``{Charming CP Violation
  and Dipole Operators from RS Flavor Anarchy},''
  \href{http://dx.doi.org/10.1007/JHEP01(2013)027}{{\em JHEP} {\bfseries 01}
  (2013) 027}, \href{http://arxiv.org/abs/1207.0474}{{\ttfamily arXiv:1207.0474
  [hep-ph]}}.

\bibitem{Keren-Zur:2012buf}
B.~Keren-Zur, P.~Lodone, M.~Nardecchia, D.~Pappadopulo, R.~Rattazzi, and
  L.~Vecchi, ``{On Partial Compositeness and the CP asymmetry in charm
  decays},'' \href{http://dx.doi.org/10.1016/j.nuclphysb.2012.10.012}{{\em
  Nucl. Phys. B} {\bfseries 867} (2013) 394--428},
  \href{http://arxiv.org/abs/1205.5803}{{\ttfamily arXiv:1205.5803 [hep-ph]}}.

\bibitem{Mannel:2012hb}
T.~Mannel and N.~Uraltsev, ``{Charm CP Violation and the Electric Dipole
  Moments from the Charm Scale},''
  \href{http://dx.doi.org/10.1007/JHEP03(2013)064}{{\em JHEP} {\bfseries 03}
  (2013) 064}, \href{http://arxiv.org/abs/1205.0233}{{\ttfamily arXiv:1205.0233
  [hep-ph]}}.

\bibitem{Fajfer:2012nr}
S.~Fajfer and N.~Ko\v{s}nik, ``{Resonance catalyzed CP asymmetries in $ D \to P
  \ell^+ \ell^- $},'' \href{http://dx.doi.org/10.1103/PhysRevD.87.054026}{{\em
  Phys. Rev. D} {\bfseries 87} no.~5, (2013) 054026},
  \href{http://arxiv.org/abs/1208.0759}{{\ttfamily arXiv:1208.0759 [hep-ph]}}.

\bibitem{Cappiello:2012vg}
L.~Cappiello, O.~Cata, and G.~D'Ambrosio, ``{Standard Model prediction and new
  physics tests for $D^0 \to h^+ h^- \ell^+ \ell^- (h=\pi,K: \ell=e,\mu)$},''
  \href{http://dx.doi.org/10.1007/JHEP04(2013)135}{{\em JHEP} {\bfseries 04}
  (2013) 135}, \href{http://arxiv.org/abs/1209.4235}{{\ttfamily arXiv:1209.4235
  [hep-ph]}}.

\bibitem{Janowski:2021yvz}
T.~Janowski, B.~Pullin, and R.~Zwicky, ``{Charged and Neutral $\bar B_{u,d,s}
  \to \gamma$ Form Factors from Light Cone Sum Rules at NLO},''
  \href{http://arxiv.org/abs/2106.13616}{{\ttfamily arXiv:2106.13616
  [hep-ph]}}.

\bibitem{Adolph:2018hde}
N.~Adolph, G.~Hiller, and A.~Tayduganov, ``{Testing the standard model with
  $D_{(s)} \to K_1 (\to K\pi\pi) \gamma$ decays},''
  \href{http://dx.doi.org/10.1103/PhysRevD.99.075023}{{\em Phys. Rev. D}
  {\bfseries 99} no.~7, (2019) 075023},
  \href{http://arxiv.org/abs/1812.04679}{{\ttfamily arXiv:1812.04679
  [hep-ph]}}.

\bibitem{ParticleDataGroup:2020ssz}
{\bfseries Particle Data Group} Collaboration, P.~A. Zyla {\em et~al.},
  ``{Review of Particle Physics},''
  \href{http://dx.doi.org/10.1093/ptep/ptaa104}{{\em PTEP} {\bfseries 2020}
  no.~8, (2020) 083C01}.

\bibitem{Buras:1994ij}
A.~J. Buras, ``{QCD factors a1 and a2 beyond leading logarithms versus
  factorization in nonleptonic heavy meson decays},''
  \href{http://dx.doi.org/10.1016/0550-3213(94)00482-T}{{\em Nucl. Phys. B}
  {\bfseries 434} (1995) 606--618},
  \href{http://arxiv.org/abs/hep-ph/9409309}{{\ttfamily arXiv:hep-ph/9409309}}.

\bibitem{Khodjamirian:1995uc}
A.~Khodjamirian, G.~Stoll, and D.~Wyler, ``{Calculation of long distance
  effects in exclusive weak radiative decays of B meson},''
  \href{http://dx.doi.org/10.1016/0370-2693(95)00972-N}{{\em Phys. Lett. B}
  {\bfseries 358} (1995) 129--138},
  \href{http://arxiv.org/abs/hep-ph/9506242}{{\ttfamily arXiv:hep-ph/9506242}}.

\bibitem{Dimou:2012un}
M.~Dimou, J.~Lyon, and R.~Zwicky, ``{Exclusive Chromomagnetism in
  heavy-to-light FCNCs},''
  \href{http://dx.doi.org/10.1103/PhysRevD.87.074008}{{\em Phys. Rev. D}
  {\bfseries 87} no.~7, (2013) 074008},
  \href{http://arxiv.org/abs/1212.2242}{{\ttfamily arXiv:1212.2242 [hep-ph]}}.

\bibitem{Burdman:1995te}
G.~Burdman, E.~Golowich, J.~L. Hewett, and S.~Pakvasa, ``{Radiative weak decays
  of charm mesons},'' \href{http://dx.doi.org/10.1103/PhysRevD.52.6383}{{\em
  Phys. Rev. D} {\bfseries 52} (1995) 6383--6399},
  \href{http://arxiv.org/abs/hep-ph/9502329}{{\ttfamily arXiv:hep-ph/9502329}}.

\bibitem{Bharucha:2015bzk}
A.~Bharucha, D.~M. Straub, and R.~Zwicky, ``{$B\to V\ell^+\ell^-$ in the
  Standard Model from light-cone sum rules},''
  \href{http://dx.doi.org/10.1007/JHEP08(2016)098}{{\em JHEP} {\bfseries 08}
  (2016) 098}, \href{http://arxiv.org/abs/1503.05534}{{\ttfamily
  arXiv:1503.05534 [hep-ph]}}.

\bibitem{Muheim:2008vu}
F.~Muheim, Y.~Xie, and R.~Zwicky, ``{Exploiting the width difference in $B_s
  \to \phi \gamma$},''
  \href{http://dx.doi.org/10.1016/j.physletb.2008.05.032}{{\em Phys. Lett. B}
  {\bfseries 664} (2008) 174--179},
  \href{http://arxiv.org/abs/0802.0876}{{\ttfamily arXiv:0802.0876 [hep-ph]}}.

\bibitem{Gratrex:2018gmm}
J.~Gratrex and R.~Zwicky, ``{Parity Doubling as a Tool for Right-handed Current
  Searches},'' \href{http://dx.doi.org/10.1007/JHEP08(2018)178}{{\em JHEP}
  {\bfseries 08} (2018) 178}, \href{http://arxiv.org/abs/1804.09006}{{\ttfamily
  arXiv:1804.09006 [hep-ph]}}.

\bibitem{Atwood:1997zr}
D.~Atwood, M.~Gronau, and A.~Soni, ``{Mixing induced CP asymmetries in
  radiative B decays in and beyond the standard model},''
  \href{http://dx.doi.org/10.1103/PhysRevLett.79.185}{{\em Phys. Rev. Lett.}
  {\bfseries 79} (1997) 185--188},
  \href{http://arxiv.org/abs/hep-ph/9704272}{{\ttfamily arXiv:hep-ph/9704272}}.

\bibitem{Greub:1996wn}
C.~Greub, T.~Hurth, M.~Misiak, and D.~Wyler, ``{The c ---\ensuremath{>} u gamma
  contribution to weak radiative charm decay},''
  \href{http://dx.doi.org/10.1016/0370-2693(96)00694-6}{{\em Phys. Lett. B}
  {\bfseries 382} (1996) 415--420},
  \href{http://arxiv.org/abs/hep-ph/9603417}{{\ttfamily arXiv:hep-ph/9603417}}.

\bibitem{Ali:1995uy}
A.~Ali and V.~M. Braun, ``{Estimates of the weak annihilation contributions to
  the decays B ---\ensuremath{>} rho + gamma and B ---\ensuremath{>} omega +
  gamma},'' \href{http://dx.doi.org/10.1016/0370-2693(95)01087-7}{{\em Phys.
  Lett. B} {\bfseries 359} (1995) 223--235},
  \href{http://arxiv.org/abs/hep-ph/9506248}{{\ttfamily arXiv:hep-ph/9506248}}.

\bibitem{Lyon:2013gba}
J.~Lyon and R.~Zwicky, ``{Isospin asymmetries in $B\to(K^*,\rho)\gamma/l^+l^-$
  and $B\to Kl^+l^-$ in and beyond the standard model},''
  \href{http://dx.doi.org/10.1103/PhysRevD.88.094004}{{\em Phys. Rev. D}
  {\bfseries 88} no.~9, (2013) 094004},
  \href{http://arxiv.org/abs/1305.4797}{{\ttfamily arXiv:1305.4797 [hep-ph]}}.

\bibitem{Ball:2006eu}
P.~Ball, G.~W. Jones, and R.~Zwicky, ``{$B \to V \gamma$ beyond QCD
  factorisation},'' \href{http://dx.doi.org/10.1103/PhysRevD.75.054004}{{\em
  Phys. Rev. D} {\bfseries 75} (2007) 054004},
  \href{http://arxiv.org/abs/hep-ph/0612081}{{\ttfamily arXiv:hep-ph/0612081}}.

\bibitem{Belle:2016mtj}
{\bfseries Belle} Collaboration, A.~Abdesselam {\em et~al.}, ``{Observation of
  $D^0\to \rho^0\gamma$ and search for $CP$ violation in radiative charm
  decays},'' \href{http://dx.doi.org/10.1103/PhysRevLett.118.051801}{{\em Phys.
  Rev. Lett.} {\bfseries 118} no.~5, (2017) 051801},
  \href{http://arxiv.org/abs/1603.03257}{{\ttfamily arXiv:1603.03257
  [hep-ex]}}.

\bibitem{Burdman:2003rs}
G.~Burdman and I.~Shipsey, ``{$D^0$ - $\bar{D}^0$ mixing and rare charm
  decays},''
  \href{http://dx.doi.org/10.1146/annurev.nucl.53.041002.110348}{{\em Ann. Rev.
  Nucl. Part. Sci.} {\bfseries 53} (2003) 431--499},
  \href{http://arxiv.org/abs/hep-ph/0310076}{{\ttfamily arXiv:hep-ph/0310076}}.

\bibitem{Donoghue:1992dd}
J.~F. Donoghue, E.~Golowich, and B.~R. Holstein,
  \href{http://dx.doi.org/10.1017/CBO9780511524370}{{\em {Dynamics of the
  standard model}}}, vol.~2.
\newblock CUP, 2014.

\bibitem{Paul:2011ar}
A.~Paul, I.~I. Bigi, and S.~Recksiegel, ``{On $D\to X_u l^+ l^-$ within the
  Standard Model and Frameworks like the Littlest Higgs Model with T Parity},''
  \href{http://dx.doi.org/10.1103/PhysRevD.83.114006}{{\em Phys. Rev. D}
  {\bfseries 83} (2011) 114006},
  \href{http://arxiv.org/abs/1101.6053}{{\ttfamily arXiv:1101.6053 [hep-ph]}}.

\end{thebibliography}\endgroup

\end{document}